\documentclass[english,aps,preprint,nofootinbib]{revtex4}
\pdfoutput=1
\usepackage[latin9]{inputenc}
\usepackage{amsmath}
\usepackage{subfig}
\usepackage{color}
\usepackage{mathrsfs}
\usepackage{indentfirst}
\usepackage{color}
\usepackage{hyperref}
\usepackage{graphicx}
\usepackage{babel}
\makeatletter

\providecommand{\LyX}{L\kern-.1667em\lower.25em\hbox{Y}\kern-.125emX\@}
\DeclareRobustCommand*{\lyxarrow}{%
\@ifstar
{\leavevmode\,$\triangleleft$\,\allowbreak}
{\leavevmode\,$\triangleright$\,\allowbreak}}

\@ifundefined{textcolor}{}
{
 \definecolor{BLACK}{gray}{0}
 \definecolor{WHITE}{gray}{1}
 \definecolor{RED}{rgb}{1,0,0}
 \definecolor{GREEN}{rgb}{0,1,0}
 \definecolor{BLUE}{rgb}{0,0,1}
 \definecolor{CYAN}{cmyk}{1,0,0,0}
 \definecolor{MAGENTA}{cmyk}{0,1,0,0}
 \definecolor{YELLOW}{cmyk}{0,0,1,0}
}

\makeatother

\begin{document}

\preprint{arXiv:1506.08276}

\title{The Quasi-normal Modes of Charged Scalar Fields in Kerr-Newman black hole and Its Geometric Interpretation}

\author{Peng Zhao}
\email{zhaopeng113@mails.ucas.ac.cn}
\affiliation{ School of Physics, University of Chinese Academy of Sciences, Beijing 100049, China }

\author{Yu Tian}
\email{ytian@ucas.ac.cn}
\affiliation{School of Physics, University of Chinese Academy of Sciences, Beijing 100049, China}
\affiliation{State Key Laboratory of Theoretical Physics, Institute of Theoretical Physics, Chinese Academy of Sciences, Beijing 100190, China}

\author{Xiaoning Wu}
\email{wuxn@amss.ac.cn}
\affiliation{Institute of Mathematics, Academy of Mathematics and System Science,Chinese Academy of Sciences, Beijing 100190, China}
\affiliation{State Key Laboratory of Theoretical Physics, Institute of Theoretical Physics, Chinese Academy of Sciences, Beijing 100190, China}

\author{Zhao-Yong Sun}
\email{sunzhaoyong11@mails.ucas.ac.cn}
\affiliation{School of Physics, University of Chinese Academy of Sciences, Beijing 100049, China }

\begin{abstract}
It is well-known that there is a geometric correspondence between high-frequency quasi-normal modes (QNMs) and null geodesics (spherical photon orbits). In this paper, we generalize such correspondence to charged scalar field in Kerr-Newman space-time. In our case, the particle and black hole are all charged, so one should consider non-geodesic orbits. Using the WKB approximation, we find that the real part of quasi-normal frequency corresponds to the orbits frequency, the imaginary part of the frequency corresponds to the Lyapunov exponent of these orbits and the eigenvalue of angular equation corresponds to carter constant. From the properties of the imaginary part of quasi-normal frequency of charged massless scalar field, we can still find that the QNMs of charged massless scalar field possess the zero damping modes in extreme Kerr-Newman spacetime under certain condition which has been fixed in this paper.
\end{abstract}
\maketitle

\tableofcontents
\newpage

\section{Introduction}
In 1970, Vishveshwara found that there exist complex frequency modes in perturbations of Schwarzschild black holes\cite{C.V.Vishveshwara}, which are called quasi-normal modes (QNMs). QNMs play an important role in the physics of black holes\cite{Press}, where many works have been done\cite{ChandraDetweiler1975,Leaver,Hod2010,Hod2012,Hod2015,Richartz}. Various aspects of QNMs continue to be uncovered in recent years. Actually, there have been some nice reviews of this subject, describing the methods to calculate QNMs and introducing the developments of QNMs:
Ref.\cite{Ferrari2008} shows some properties of eigenfrequencies and potential applications of QNMs to gravitational wave asteroseismology;
Ref.\cite{Kokkotas} introduces some analysis about the application of QNMs to relativistic stars and the detection of QNMs; Ref.\cite{Berti2009} discusses the applications of QNMs in the AdS/CFT duality and to astrophysical black holes; Ref.\cite{Konoplya} discusses the gravitational instabilities in higher than four dimensions and the AdS/CFT interpretation of QNMs.

The semi-classical interpretation of QNMs is particularly interesting. In 1984, Ferrari and Mashhood\cite{Ferrari1984} found that the quasi-normal frequency in Schwarzschild black hole can be written as
\begin{align}
\omega\approx\left(l+\frac{1}{2}\right)\Omega-i\left(n+\frac{1}{2}\right)\gamma_L\nonumber
\end{align}
when $l\gg1$. Here $\Omega$ is the frequency of circular photon orbits and $\gamma_L$ is the Lyapunov exponent of the orbits which describe how quickly the cross section of the null geodesic congruence changed under infinitesimal radial perturbations. In the eikonal limit, the authors of \cite{schutz} studied the QNMs by the WKB approximation and obtained the expression of the imaginary part of the quasi-normal frequency. The relationship between the QNMs and the circular null geodesics was then studied in \cite{Cardoso2009}, which showed that the QNMs of black holes in any dimensions are determined by the parameters of the circular null geodesics. Recently, the authors of \cite{H.Yang} made a detailed study of the geometric interpretation of the QNMs of massless scalar field in general Kerr black holes by the WKB approximation.

Usually, people may think that the charge effect is unimportant since most celestial bodies are electroneutral. But things are not always like that. Recently, some super-Chandrasekhar white dwarfs have been found\cite{Howell,Scalzo} and this phenomenon can be explained by the charge effect\cite{Olson,Herrera,Xiangdong Zhang}. Even if the particle is uncharged, the behaviors of its trajectories near the singularity of Schwarzschild and Reissner-Nordstrom (RN) black holes are quite different. For the RN black hole, the singularity is untouchable by geodesics, but this phenomenon does not occur for the Schwarzschild black hole. Furthermore, when the matter fields or perturbations in Kerr-Newman (KN) black hole are charged, the interaction between matter fields and electromagnetic field in the KN black hole will have significant influence on the QNMs, which has been studied in, e.g., \cite{Berti,Hod2008,Kokkotas2011,Konoplya2013,Mark2014}.

So it is interesting to study the geometric interpretation of QNMs of the charged scalar field in the KN black hole, which is the main motivation of this paper. In order to achieve this geometric interpretation, we also need the behavior of particle motion in the KN spacetime, which can be found in \cite{Hackmann2013,Hackmann2015,Ulbricht}, and generalize the method in \cite{H.Yang} to explore the geometric correspondence between the QNMs and spherical orbits of charged particles for the KN black hole, where special attention is paid to the non-geodesic nature of these orbits. It is shown that, noting the $U(1)$ gauge invariance of physical quantities, the QNMs have perfect geometric interpretation by means of the non-geodesic world line congruence. The influence of charge of the particle is also addressed.

We organize the rest of the paper as follows. In Sec.~\ref{particle} we will investigate the Hamilton-Jacobi formalism of particle motion and study the geometric correspondence of QNMs. We can verify that those relationships\cite{H.Yang} also hold: in the leading order, the real part of quasi-normal frequency corresponds to the energy $E$, the azimuthal quantum number corresponds to the angular momentum $L_z$, and the real part of angular eigenvalue corresponds to the Carter constant $\mathcal{Q}$; in next-to-leading order, the imaginary part of quasi-normal frequency corresponds to the Lyapunov exponent $\gamma_L$ of radial motion, and the next-to-leading order correction of the angular eigenvalue corresponds to the imaginary part of the Carter constant. In Sec.~\ref{further} we obtain the geometric correspondence by some specific example and study the influence of charge. We also study and find the condition of the zero damping modes in extreme KN black hole. We can observe that the imaginary part of the quasi-normal frequency will approach to zero under this condition. In Sec.~\ref{conclusion} we conclude with some discussion.

\section{Quasi-normal modes in Kerr-Newman spacetime}\label{qnms1}
We consider the charged massive scalar field $u$ in KN spacetime in Boyer-Lindquist coordinate system:
\begin{align}
ds^2=&-\left(1-\frac{2Mr-Q^2}{\rho^2}\right)dt^2+\frac{\rho^2}{\Delta}dr^2+\rho^2d\theta^2+\frac{1}{\rho^2}\Big(\left(r^2+a^2\right)^2 \\ \nonumber
&-\Delta a^2\sin^2\theta\Big)\sin^2\theta d\phi^2-\frac{2a}{\rho^2}\left(2Mr-Q^2\right)\sin^2\theta dtd\phi.
\end{align}
with $M$ is the mass of black hole, $aM$ is the angular momentum of black hole, $Q$ is the black hole charge,  $\rho^2=r^2+a^2\cos^2\theta$, $\Delta=r^2-2Mr+a^2+Q^2$ and the four potential $A_a=-\frac{Qr}{\rho^2}[(dt)_a-a\sin^2\theta (d\phi)_a]$.
The equation of motion of $u$ is the Klein-Gordon(KG) equation with charge in curved spacetime:
\begin{equation}
g^{ab}(\nabla_a-iqA_a)(\nabla_b-iqA_b)u-\mu^2_\star u=0.
\end{equation}
with $q$ is the charge of the scalar field, $\mu_\star $ is the mass of the field. We can obtain that $\nabla_aA^a=0$, $A^aA_a=-\frac{Q^2r^2}{\rho^2\Delta}$, $A^0=\frac{Qr}{\rho^2\Delta}(r^2+a^2)$, $A^3=\frac{aQr}{\rho^2\Delta}$ under this gauge.
 We can separate the scalar field as:
\begin{equation}
u(t,r,\theta,\phi)=\sum_{l,m}\int e^{-i\omega t}e^{im\phi}R(r)u_\theta(\theta)\,d\omega.
\end{equation}
and obtain the radial equation and the angular equation as:
\begin{equation}\label{radial1}
\frac{d}{dr}\left(\Delta\frac{dR}{dr}\right)+\left(\frac{\left(K+qQr\right)^2}{\Delta}-\left(A_{lm}+a^2\omega^2-2am\omega+\mu^2_\star r^2\right)\right)R=0;
\end{equation}
\begin{equation}\label{theta1}
\frac{1}{\sin\theta}\frac{d}{d\theta}\left(\sin\theta\frac{d}{d\theta}u_\theta\right)+\left(\left(a^2\omega^2-a^2\mu_\star^2\right)\cos^2\theta-\frac{m^2}{\sin^2\theta}+A_{lm}\right)u_\theta=0
\end{equation}
with $K=-\omega(r^2+a^2)+am$, $A_{lm}$ is the angular eigenvalue of the angular equation and depend on quantum number $l$ and $m$. This is the Teukolsky equation(\cite{Teukolsky}) with spin $s$ and source are zero.
\subsection{The angular eigenvalue equation}
Under the transformation
\begin{equation}
x=\ln\left(\tan\frac{\theta}{2}\right),
\end{equation}
namely $dx=\frac{d\theta}{\sin\theta}$, the angular eigenvalue equation (\ref{theta1}) can be written as
\begin{equation}\label{theta2}
\frac{d^2u_\theta}{dx^2}+V^\theta u_\theta=0.
\end{equation}
with $V^\theta=a^2(\omega^2-\mu_\star^2)\cos^2\theta \sin^2\theta-m^2+A_{lm}\sin^2\theta$.
We can see that the influence of mass can be reflected in $\omega^2-\mu_\star^2$. {We assume that $m\ne 0$ and the condition $0<\mu_\star<|\omega|$ holds in the following discussion. Since} the quasi-normal frequency is complex with $\omega=\omega_R-i\omega_I$, the angular value $A_{lm}$ is also complex:
\begin{equation}
A_{lm}=A^R_{lm}+iA_{lm}^I.
\end{equation}
We can use the WKB analysis to deal with this equation, the approximate solution is
\begin{align}
&u_\theta=\frac{c_-\exp\left(-\int_x^{x_-}\sqrt{-V^{\theta}\left(x^{\prime}\right)}\,dx^{\prime}\right)}{\mid V^{\theta}(x^{\prime})\mid^\frac{1}{4}},\qquad x<x_- \\
&u_\theta=\frac{a_+\exp\left(i\int_{x_-}^{x}\sqrt{V^{\theta}\left(x^{\prime}\right)}\,dx^{\prime}\right)+a_-\exp\left(-i\int_{x_-}^{x}\sqrt{V^{\theta}\left(x^{\prime}\right)}\,dx^{\prime}\right)}{\mid V^{\theta}(x^{\prime})\mid^\frac{1}{4}},\qquad x_-<x<x_+ \\
&u_\theta=\frac{c_+\exp\left(-\int_{x_+}^{x}\sqrt{-V^{\theta}\left(x^{\prime}\right)}\,dx^{\prime}\right)}{\mid V^{\theta}(x^{\prime})\mid^\frac{1}{4}},\qquad x>x_+
\end{align}
where $x_-$ and $x_+$ is the turning point of $V^\theta(x)$, i.e. $V^\theta(x)$ is equal to zero at these two points. The wave propagates in the region $x_-<x<x_+$ and decay to zero when $x\to\pm\infty$.

The Bohr-Sommerfeld condition is
\begin{equation}\label{BScondition1}
\int_{\theta_-}^{\theta_+} d\theta\, \sqrt{a^2(\omega^2_R-\mu_\star^2)\cos^2\theta-\frac{m^2}{\sin^2\theta}+A^R_{lm}}=(L-\vert m\vert)\pi.
\end{equation}
with $L=l+\frac{1}{2}$. We define $\mu\equiv{m}/{L}$, $\alpha_R(a,\mu)\equiv{A^R_{lm}}/{L^2}$, $m_*\equiv{\mu_*}/{L}$, $\Omega_R(a,\mu)\equiv{\omega_R}/{L}$.

We can treat $a\sqrt{\Omega^2_R-m^2_\star}$ as a small parameter and expand this integration to obtain
\begin{equation}
\alpha_R\approx1-\frac{a^2(\omega^2_R-\mu^2_\star)}{2L^2}(1-\mu^2).
\end{equation}
Using the facts that $\omega_R \sim O(l)$, $\omega_I \sim O(1)$, 
$m \sim O(l)$ (see \cite{H.Yang}), we obtain that $\omega_R\gg\omega_I$. Then we can use the perturbation theory of eigenvalue equation\cite{H.Yang}, which leads to the next-to-leading-order correction of $A_{lm}$:
\begin{align}
A^I_{lm}=-2a^2\omega_R\omega_I\langle\cos^2\theta\rangle=a\omega_I\left[\frac{\partial A^R_{lm}(z)}{\partial z}\right]_{z=a\omega_R}.
\end{align}
Finally we can obtain the approximate expression of $A_{lm}$ in eikonal limit ($l\gg1$) as
\begin{equation}
A_{lm}\approx l(l+1)-\frac{a^2(\omega^2-\mu^2_\star)}{2}\left[1-\frac{m^2}{l(l+1)}\right].
\end{equation}

\subsection{The radial eigenvalue equation}
We define
\begin{equation}
u_r=\sqrt{r^2+a^2}R,\qquad\frac{d}{dr_\star}=\frac{\Delta}{r^2+a^2}\frac{d}{dr},
\end{equation}
and
\begin{equation}
V^r=-\frac{\Delta}{(r^2+a^2)^2}\left[\lambda_{lm}+\mu_\star^2r^2+\frac{\Delta+2r\left(r-M\right)}{r^2+a^2}-\frac{3\Delta r^2}{\left(r^2+a^2\right)^2}\right]+\left(\omega-\frac{ma+qQr}{r^2+a^2}\right)^2
\end{equation}
with $\lambda_{lm}=A_{lm}+a^2\omega^2-2am\omega$. Then the radial eq.(\ref{radial1}) can be transformed as
\begin{equation}\label{radial2}
\frac{d^2u_r}{dr^2_\star}+V^ru_r=0.
\end{equation}
In the condition of eikonal limit ($l\gg1$), we can neglect the terms $\frac{\Delta+2r(r-M)}{r^2+a^2}-\frac{3\Delta r^2}{(r^2+a^2)^2}$ in the expression of the effective potential $V^r$, we rewrite the effective potential as
\begin{equation}
V^r=\frac{[\omega(r^2+a^2)-ma-qQr]^2-\Delta (A_{lm}+a^2\omega^2-2ma\omega+\mu_\star^2r^2)}{(r^2+a^2)^2}.
\end{equation}
Under this transformation, when $r\rightarrow +\infty$, we have $r_\star\rightarrow +\infty$ and the effective potential $V^r\to\omega^2-\mu_\star^2$; when $r\rightarrow r_+$, we have $r_\star\rightarrow -\infty$ and $V^r\to(\omega-\frac{ma+qQr_+}{r_+^2+a^2})^2$. In order to obtain QNMs, we require that the wave is outgoing at $r_\star\rightarrow +\infty$ and ingoing at  $r_\star\rightarrow -\infty$ . By comparing handling with the stationary Schrodinger equation, we know that boundary condition require that there is a point $r^\star_0$ such that the effective potential $V^r$ is zero here. We can expand effective potential $V^r$ here as
\begin{align}
V^r(r_\star)=V_0+V^\prime_0(r_\star-r^\star_0)+\frac{V^{\prime\prime}_0}{2}(r_\star-r^\star_0)^2+O((r_\star-r^\star_0)^3),
\end{align}
with $V^\prime_0=0$, $V_0=V^r(r^\star_0)$, $V^\prime_0=\frac{\partial V^r}{\partial r_\star}\mid_{r^\star_0}$, $V^{\prime\prime}_0=\frac{\partial^2V^r}{\partial r_\star^2}\mid_{r^\star_0}$. Taking this Taylor expansion into radial equation and using the result of Schutz and Will\cite{schutz} as well as boundary conditions, we can obtain
\begin{align}\label{xb}
n+\frac{1}{2}=\frac{iV_0}{(2V^{\prime\prime}_0)^{\frac{1}{2}}}.
\end{align}

Due to the spirit of QNMs, we write $\omega=\omega_R-i\omega_I$, where the imaginary part is a small parameter in contrast to the real part. When we take this into (\ref{xb}) and $V^\prime_0=0$, and expand with respect to $\omega$, we obtain
\begin{align}
&V^r(r^\star_0,\omega_R)=0=\frac{\partial V^r}{\partial r_\star}\mid_{r^\star_0} \\
&\omega_I=(n+\frac{1}{2})\frac{\sqrt{2(\frac{d^2 V_r}{dr_\star^2})_{r^\star_0,\omega_R}}}{(\frac{\partial V_r}{\partial \omega})_{r^\star_0,\omega_R}}
\end{align}
In other words, there is a radius $r_0$ such that
\begin{align}\label{ddsb}
V^r(r_0,\omega_R)=0=\frac{\partial V^r}{\partial r}(r_0,\omega_R) ;
\end{align}
\begin{align}\label{jsxb}
\omega_I=(n+\frac{1}{2})\frac{\sqrt{2(\frac{d^2 V_r}{dr_\star^2})_{r_0,\omega_R}}}{(\frac{\partial V_r}{\partial \omega})_{r_0,\omega_R}} .
\end{align}
Then we can use (\ref{ddsb}) to calculate the position of apex of effective potential and the real part of frequency, and obtain the imaginary part of frequency.

\section{QNMs and the motion of particle in KN spacetime}\label{particle}

\subsection{Geometric optics in KN spacetime}
The massive charged scalar field in curved spacetime satisfy the KG equation
\begin{equation}\label{KGeq1}
g^{ab}(\nabla_a-iqA_a)(\nabla_b-iqA_b)u-\mu^2_\star u=0.
\end{equation}
We assume that the wave function can be expressed as
\begin{equation}\label{formu1}
u=Ae^{i\Phi}.
\end{equation}
When we taking this formula (\ref{formu1}) into the KG equation.(\ref{KGeq1}) and define $k_\mu=\partial_\mu\Phi$, the leading order equation is
\begin{equation}\label{leadingorder1}
g^{\mu\nu}(k_\mu-qA_\mu)(k_\nu-qA_\nu)+\mu^2_\star=0,
\end{equation}
and the next-to-leading order equation is
\begin{equation}\label{nextleading1}
2(k^\mu-qA^\mu)\partial_\mu\ln A+\nabla_\mu(k^\mu-qA^\mu)=0.
\end{equation}
We know that the mechanical moment $k_\mu-qA_\mu$ is how wave propagate in curved background from leading order \ref{leadingorder1} and we note $U^{\mu}=k^{\mu}-qA^{\mu}$ to represents the four-velocity. The electromagnetic field makes particle's worldline no longer geodesic and that the phase will change along the particle's worldline:
\begin{equation}
(U^\mu-qA^\mu)\partial_\mu\Phi=-\mu^2_\star+\frac{q^2Q^2r^2}{\rho^2\Delta}.
\end{equation}

Furthermore, we can obtain $U^{\mu}\nabla_{\mu}U^{\rho}=qF^{\rho}_{\phantom{\rho}\mu}U^{\mu}$ from the leading order, that is the equation of motion of a charged massive particle in a curved spacetime with a electromagnetic field. It is interesting to know that the mechanical moment $U^{\mu}$ is different from the canonical momentum $k_\mu$. As we all know that the electromagnetic field is gauge field and possesses the $U(1)$ gauge invariance, when we do the gauge transformation
\begin{equation}
A_\mu\to A_{\mu}^{\prime}=A_\mu+\partial_\mu\chi.
\end{equation}
then the phase of wave function will changed correspondingly:
\begin{equation}
Ae^{i\Phi}\to Ae^{i\Phi+i\chi}\qquad\Phi\to \Phi^{\prime}=\Phi+\chi,
\end{equation}
and the canonical momentum will change like the following:
\begin{equation}
k_\mu\to k_\mu^{\prime}=k_\mu+\partial_\mu\chi.
\end{equation}
We can obtain that the wave front will change under gauge transformation but the four velocity will invariant $U_{\mu}=U_{\mu}^{\prime}$.

The next-to-leading order equation (\ref{nextleading1}) gives rise to
\begin{align}
\nabla_\mu U^\mu=-2U^\mu\partial_\mu\ln A.
\end{align}
Similar to \cite{H.Yang}, this equation is related to the fluid expansion equation
\begin{align}
\nabla_\mu U^\mu=U^\mu\partial_\mu\ln\mathscr{A}
\end{align}
with $\mathscr{A}$ represents the area of cross section of the congruence. So the right hand side of above equation shows the change rate of cross section along the worldline. After comparing these two equations we have
\begin{align}
U^\mu\partial_\mu\ln(\mathscr{A}^{\frac{1}{2}}A)=0,
\end{align}
so
\begin{align}\label{mjzf}
A\propto\mathscr{A}^{-\frac{1}{2}}.
\end{align}
It is well-known that the square of the length of amplitude represents the probability of the appearance of particle, this result tells us that the more the area of cross section of congruence, the less the probability of the appearance of particle.

\subsection{Equation of motion for particles in KN spacetime}
We will use the Hamilton-Jacobi formalism to study the motion of charged massive particle:
\begin{equation}
g^{\mu\nu}(\partial_\mu S-qA_\mu)(\partial_\nu S-qA_\nu)+\mu^2_\star =0.
\end{equation}
with $S$ represents the Hamilton principal function. With two Killing vector fields $\big(\frac{\partial}{\partial t}\big)^a$ and $\big(\frac{\partial}{\partial\phi}\big)^a$, S can be written into the following form:
\begin{equation}
S=\frac{\mu^2_\star\tau}{2}-Et+L_z\phi+S_r(r)+S_\theta(\theta).
\end{equation}
where $E$ represents the total energy of particle with gravitational as well as electromagnetic energy, $L_z$ represents the z-directed specific angular momentum. The property of timelike and axial Killing vectors ensure us that $E$ and $L_z$ are conserved constants. And $\tau$ is the proper time of particle. The angular and radial equations are
\begin{equation}
(S_\theta^\prime)^2=\mathcal{Q}-\cos^2\theta\left(\frac{L_z^2}{\sin^2\theta}-a^2E^2+a^2\mu^2_\star\right),
\end{equation}
\begin{equation}
\Delta^2(S_r^\prime)^2=\left(E\left(r^2+a^2\right)-aL_z-qQr\right)^2-\Delta\left(\left(L_z-aE\right)^2+\mathcal{Q}+\mu^2_\star r^2\right),
\end{equation}
where $S_\theta^\prime$ represents derivative with respect to $\theta$ and $S_r^\prime$ with respect to $r$, $\mathcal{Q}$ is Carter constant which is the third conserved quantity. We can see how the charge and mass influence the radial equation ($qQr$ and $\mu^2_\star r^2$), but there is no charge influence in angular equation. We define
\begin{equation}
\mathcal{R}(r)=\left(E\left(r^2+a^2\right)-aL_z-qQr\right)^2-\Delta\left(\left(L_z-aE\right)^2+\mathcal{Q}+\mu^2_\star r^2\right),
\end{equation}
\begin{equation}
\Theta(\theta)=\mathcal{Q}-\cos^2\theta\left(\frac{L_z^2}{\sin^2\theta}-a^2E^2+a^2\mu^2_\star\right).
\end{equation}
Then we have
\begin{equation}
S_r(r)=\int^r\frac{\sqrt{\mathcal{R}}}{\Delta}\,dr
\end{equation}
\begin{displaymath}
S_\theta(\theta)=\int^\theta\sqrt{\Theta}\,d\theta.
\end{displaymath}
So the Hamilton principal function has the form
\begin{equation}
S=\frac{\mu^2_\star\tau}{2}-Et+L_z\phi+\int^r\frac{\sqrt{\mathcal{R}}}{\Delta}\,dr+\int^\theta\sqrt{\Theta}\,d\theta.
\end{equation}

From $\frac{\partial S}{\partial\mu^2_\star}=0$ and $\frac{\partial S}{\partial\mathcal{Q}}=0$, we can obtain
\begin{equation}
\int^r\frac{r^2}{\sqrt{\mathcal{R}}}\,dr+\int^\theta\frac{a^2\cos^2\theta}{\sqrt{\Theta}}\,d\theta=\tau,
\end{equation}
\begin{equation}
\int^r\frac{1}{\sqrt{\mathcal{R}}}\,dr=\int^\theta\frac{1}{\sqrt{\Theta}}\,d\theta.
\end{equation}
Taking derivatives of these two equations with respect to $\lambda$:
\begin{equation}
\frac{d}{d\lambda}=\rho^2\frac{d}{d\tau},
\end{equation}
we have
\begin{equation}\label{radial4}
\frac{r^2}{\sqrt{\mathcal{R}}}\frac{dr}{d\lambda}+\frac{a^2\cos^2\theta}{\sqrt{\Theta}}\frac{d\theta}{d\lambda}=\rho^2,
\end{equation}
\begin{equation}\label{radial3}
\frac{1}{\sqrt{\mathcal{R}}}\frac{dr}{d\lambda}=\frac{1}{\sqrt{\Theta}}\frac{d\theta}{d\lambda}.
\end{equation}
We assume the second eq.(\ref{radial3}) is equal to $f(\lambda)$ and take it into the first equation.(\ref{radial4}), we can obtain that $f(\lambda)=1$. Then these two equations will be recast into
\begin{equation}
\frac{dr}{d\lambda}=\sqrt{\mathcal{R}},
\end{equation}
\begin{equation}
\frac{d\theta}{d\lambda}=\sqrt{\Theta}.
\end{equation}

From $\frac{\partial S}{\partial E}=0$ and $\frac{\partial S}{\partial L_z}=0$, we have
\begin{equation}
t=\int^r\frac{(r^2+a^2)\left(E\left(r^2+a^2\right)-aL_z-qQr\right)-a\Delta\left(aE-L_z\right)}{\Delta\sqrt{\mathcal{R}}}+\int^\theta\frac{a^2E^2\cos^2\theta}{\sqrt{\theta}},
\end{equation}
\begin{equation}
\phi+\int^r\frac{a\left(aL_z+qQr-E\left(r^2+a^2\right)\right)-\Delta(L-aE)}{\Delta\sqrt{\mathcal{R}}}-\int^\theta\frac{L_z \cos^2\theta}{\sqrt{\theta}\sin^2\theta}=0
\end{equation}
When we derivative these two equations with respect to $\lambda$ we can obtain
\begin{equation}
\frac{dt}{d\lambda}=\frac{r^2+a^2}{\Delta}\left(E\left(r^2+a^2\right)-aL_z-qQr\right)-a(aE\sin^2\theta-L_z),
\end{equation}
\begin{equation}
\frac{d\phi}{d\lambda}=-\left(aE-\frac{L_z}{\sin^2\theta}\right)+\frac{a\left(E\left(r^2+a^2\right)-aL_z-qQr\right)}{\Delta}.
\end{equation}
Based on the above two equations, the mass of the particle plays no role but the charge of the particle will have physical effects.

\subsection{Leading order correspondence with QNMs}
Based on the WKB method, the phase of the wave function corresponds to the Hamilton principal function, the leading order approximation of wave equation corresponds to Hamilton-Jacobi equation. So the wave function has the following form:
\begin{equation}
u=e^{iS}=e^{-iEt}e^{iL_z\phi}e^{\pm iS_\theta}e^{\pm iS_r}.
\end{equation}
By comparing to the preceding equations of QNMs, we can still find that the conserved particle energy $E$ corresponds to the real part of quasi-normal frequency $\omega_R$, the z-directed angular momentum $L_z$ corresponds to the azimuthal quantum number $m$. When we regard $u_\theta$ as $e^{iS_\theta}$, we can obtain $\mathcal{Q}=A_{lm}^R-m^2$. These is the same as what we get in Kerr back ground. However we cannot get the damping behaviour in leading order. In the next section we will discuss the next-to-leading order behaviour.

\subsection{Next-to-leading order correspondence with QNMs}

\subsubsection{The imaginary part of quasi-normal frequency with Lyapunov exponent of radial motion}
In last section we show that the conserved quantities of particle orbit ($E,\mathcal{Q},L_z$) correspond to the real part of parameters of a QNMs $(\omega_R,A^R_{lm},m)$. They are the leading order behaviour. The damping behaviour of QNMs will appear in the next-to-leading order. In this section we will discuss this phenomenon. By the symmetry, we can obtain that there should not be any correction in $\phi$-direction. QNMs tell us that the correction of $t$- direction will decay. Then we write the wave function as\cite{H.Yang}
\begin{equation}\label{ampltude1}
u=Ae^{iS}=e^{-\gamma t}A_r(r)A_\theta(\theta)e^{-iEt}e^{iL_z\phi}e^{\pm iS_\theta}e^{\pm iS_r}.
\end{equation}

We define $A(t,r,\theta)=e^{-\gamma t}A_r(r)A_\theta(\theta)$. Then the next-to-leading order condition \ref{nextleading1} gives rise to
\begin{equation}\label{ampltude2}
\rho^2\frac{d\ln A}{d\tau}=-\frac{1}{2}\left(\partial_r\left(\Delta\partial_rS_r\right)+\frac{1}{\sin\theta}\partial_\theta\left(\sin\theta\partial_\theta S_\theta\right)\right).
\end{equation}
with $\tau$ the affine parameter. We should note that $U^\mu\nabla_\mu=\frac{d}{d\tau}$ which is different from $k^\mu\nabla_\mu$. The transformation $\frac{d}{d\lambda}=\rho^2\frac{d}{d\tau}$ can change the left hand side of this equation as:
\begin{equation}\label{ampltude3}
\frac{d\ln A}{d\lambda}=\frac{d\ln A_r}{d\lambda}+\frac{d\ln A_\theta}{d\lambda}-\gamma\frac{dt}{d\lambda}.
\end{equation}

The expression of $\frac{dt}{d\lambda}$ can be separated into two part: one is only depend on $r$ and another on $\theta$:
\begin{equation}\label{ampltude4}
\frac{dt}{d\lambda}=\bar{\dot{t}}+\tilde{\dot{t}}.
\end{equation}
with $\bar{\dot{t}}$ is the function of $r$ and $\tilde{\dot{t}}$ is the function of $\theta$. \\

We assume that $A_r$ is a good function near $r=r_0$ or analytic,  then we have $A_r(r)\sim(r-r_0)^n$ around $r_0$ with n is a integer. Then we can obtain
\begin{equation}
\gamma=\left(n+\frac{1}{2}\right)\frac{\sqrt{\mathcal{R}_0^{\prime\prime}/2}}{\bar{\dot{t}}}=\left(n+\frac{1}{2}\right)\lim_{r\to r_0}\frac{1}{r-r_0}\frac{dr/d\lambda}{\langle dt/d\lambda\rangle_\theta}.
\end{equation}
where $\mathcal{R}_0^{\prime\prime}\equiv\mathcal{R}^{\prime\prime}(r_0)$.

The physical interpretation here is that the motion in the $\theta$-direction is independent of $r$. Then we can choose a worldline congruence whose $r$-value is a little bigger than $r_0$ but very different $\theta$. When they return their respective initial value of $\theta$, the change of radial $r$ $\Delta r$ is far less than $r_0$. And we have that motion along a period of $\theta$ :
\begin{equation}
\frac{1}{r-r_0}\frac{\Delta r}{\Delta t}\equiv \gamma_L
\end{equation}
with right side will approach to a constant.As we all know the definition of Lyapunov exponent :
\begin{align}
\lambda=\lim_{\substack{t\to\infty\\\mid\Delta X_0\mid\to 0}}\frac{1}{t}\ln\frac{\mid\Delta X(X_0,t)\mid}{\mid\Delta X_0\mid}.
\end{align}
with $\mid\Delta X_0\mid$ represents the interval between two paths at initial time in phase space, $\mid\Delta X(X_0,t)\mid$ represents the interval between these two paths after time $t$. Under these circumstances, one path is a unstable circular orbit which the initial radial position is $r_0$ and will be at $r_0$ after a period $\Delta t$ of $\theta$ motion, the initial position of another path is a little bigger than $r_0$, for example $r$, and will be at $r+\Delta r$ after $\Delta t$. By the definition of Lyapunov exponent, we have:
\begin{align}
\lambda&\approx\frac{1}{\Delta t}\ln\frac{r+\Delta r-r_0}{r-r_0}=\frac{1}{\Delta t}\ln\left(1+\frac{\Delta r}{r-r_0}\right) \\ \nonumber
&\approx\frac{1}{r-r_0}\frac{\Delta r}{\Delta t}=\gamma_L.
\end{align}
So the physical meaning of $\gamma_L$ is the Lyapunov exponent of radial motion. The comparison between the circular orbit and the orbit around it tell us that the worldline will return their initial $\theta$ position after a period of $\theta$ motion. The $\phi$-direction is Killing direction, this symmetry guarantee that the $\phi$-direction of congruence will not change. Then the area of cross-section $\mathscr{A}$ only depends on the radial direction: $\tilde{\Delta r}(\Delta t)=r+\Delta r-r_0$. Base on the definition $\tilde{\Delta r}(\Delta t)=e^{\gamma_L\Delta t}\tilde{\Delta r}$ of the Lyapunov exponent, we can obtain
\begin{align}
\mathscr{A}\propto e^{\gamma_Lt}.
\end{align}
So we have:
\begin{align}
A\propto e^{-\gamma_Lt/2}.
\end{align}

In the following we consider a worldline whose initial coordinates are $(t,r,\theta,\phi)$ and will be $(t+\Delta t,r+\Delta r,\theta,\phi+\Delta\phi)$ after a period of $\theta$. In this process, the condition $\frac{\partial S}{\partial E}=0$ and $\frac{\partial S}{\partial\mathcal{Q}}=0$ implies
\begin{equation}
\frac{\partial}{\partial E}\Big[\int^{r+\Delta r}_r\frac{\sqrt{\mathcal{R}(r^\prime)}}{\Delta(r^\prime)}dr^\prime+\Delta S_\theta\Big]=\Delta t;
\end{equation}
\begin{equation}
\frac{\partial}{\partial\mathcal{R}}\Big[\int^{r+\Delta r}_r\frac{\sqrt{\mathcal{R}(r^\prime)}}{\Delta(r^\prime)}dr^{\prime}+\Delta S_\theta\Big]=0.
\end{equation}
with:
\begin{equation}
\Delta S_\theta=2\int^{\theta_+}_{\theta_-}\sqrt{\Theta(\theta^\prime)}\,d\theta^\prime\equiv\oint\sqrt{\Theta(\theta^\prime)}\,d\theta^\prime.
\end{equation}
Due to the radial $r$ is near $r_0$ and will change little, then the integral with respect to $r^\prime$ can be changed as the product of integrand and $\Delta r$. Using the Bohr-Sommerfeld condition $\Delta S_\theta=2(L-|m|)$ and its total derivative with respect to $E$ is zero as well as $\mathcal{R}(r)\approx\frac{(r-r_0)^2}{2}\mathcal{R}^{\prime\prime}$, we can obtain\cite{H.Yang}
\begin{equation}\label{gamma}
\gamma=(n+\frac{1}{2})\frac{\sqrt{2\mathcal{R}_0^{\prime\prime}}\Delta_0}{\Big[\frac{\partial\mathcal{R}}{\partial E}+\frac{\partial\mathcal{R}}{\partial\mathcal{Q}}\big(\frac{d\mathcal{Q}}{dE}\big)_{BS}\Big]_{r_0}}
\end{equation}
where $\Delta_0=\Delta(r_0)$.

We can see that this is equivalent with the expression of imaginary part of quasi-normal frequency.
At last we can sum up what we have done: the imaginary part of massive charged quasi-normal frequency of KN black hole is respond to the Lyapunov exponent of radial motion of massive charged particle of KN spacetime. The imaginary part of frequency represents the damping behaviour of wave and the Lyapunov exponent of particle represent the expansion of the cross section of the worldline congruence around the unstable circular orbits. Based on the expression (\ref{gamma}), since $\mathcal R$ and $\mathcal Q$ both depend on $q$, the charge of the particle plays an important role in the motion.
\subsubsection{Angular amplitude correction and the imaginary part of Carter constant}
We can see that the Carter constant corresponds to the real part of angular eigenvalue:$$\mathcal{Q}=A^R_{lm}-m^2$$ from the leading order equation, and we know that $A_{lm}$ is complex. We can guess that the Carter constant can be complex and this equation can still be hold. As we can see that particle charge does not appear in $\Theta(\theta)$, so does the angular amplitude correction $A_\theta$, so charge will not affect the correction. Although $\Theta(\theta)$ depends on the particle mass, it does not have essential effect on the correction. Then the meaning of complex $A_{lm}$ can also be explained by the complex Carter constant.

From eq.(\ref{ampltude1},\ref{ampltude2},\ref{ampltude3},\ref{ampltude4}), we can obtain the correction:
\begin{equation}
A_{\theta}=\frac{\exp\left(\left(-i\gamma\right)\big[\frac{\partial}{\partial E}+\big(\frac{d\mathcal{Q}}{dE}\big)_{BS}\frac{\partial}{\partial\mathcal{Q}}\big]\left(iS_{\theta}\right)\right)}{\sqrt{\sin\theta\sqrt{\Theta}}}.
\end{equation}
All the $\mathcal{Q}$ and $E$ previous are real number and in next part we will mark them as $\mathcal{Q}_R$ and $E_R$ and let $\mathcal{Q}$ and $E$ be complex, namely $\mathcal{Q}=\mathcal{Q}_R+i\mathcal{Q}_I$ and $E=E_R+iE_I$. As we have know that $E_I=-\gamma=-\omega_I$, If we let $\mathcal{Q}_I=-\gamma\big(\frac{d\mathcal{Q}}{dE}\big)_{BS}$, and see the imaginary part as a small parameter and expand $\exp(iS_\theta)$ which the parameters are complex, then we can obtain:
\begin{eqnarray}
&&\exp(iS_\theta)=\exp\left(i\int^{\theta}_{\theta_-}\sqrt{\Theta}\,d\theta\right)\\ \nonumber
&&=\sqrt{\sin\theta\sqrt{\Theta}}A_{\theta}\exp(iS^R_{\theta})
\end{eqnarray}
where the parameters in $S^R_{\theta}$ are real\cite{H.Yang}. We can see that the $A^I_{lm}=\mathcal{Q}_I$ from the expression of imaginary part of angular eigenvalue.

In summary, the correction of wave amplitude make the conserved quantity $E$ and $\mathcal{Q}$ be complex, the imaginary part of Carter constant correspond to the next-to-leading order correction of angular amplitude, then the relationship $\mathcal{Q}=A_{lm}-m^2$ can hold for a complex $\mathcal{Q}$ and $A_{lm}$.

\section{Effects of charge on QNMs}\label{further}
\subsection{Unstable circular orbit and spherical orbit's inclination angle}
We know that the radial equation :
\begin{equation}
\frac{dr}{d\lambda}=\sqrt{\mathcal{R}}
\end{equation}
is the first integral of motion, where :
\begin{equation}
\mathcal{R}(r)=\left(E\left(r^2+a^2\right)-aL_z-qQr\right)^2-\Delta\left(\left(L_z-aE\right)^2+\mathcal{Q}\right)
\end{equation}
The equation of motion is a second order differential equation, namely:
\begin{align}
\frac{d^2r}{d\lambda^2}=\frac{d\sqrt{\mathcal{R}}}{d\lambda}\frac{d r}{d\lambda}=\frac{1}{2}\frac{d\mathcal{R}}{dr}.
\end{align}
Then the condition of unstable circular orbit means that (by \cite{Chandrasekhar})
\begin{align}
\mathcal{R}(r)=0=\mathcal{R}^{\prime}(r),
\end{align}
namely
\begin{align}
&-\left(\Delta-a^2\right)\xi^2-2a[(2M-\sigma Q)r-Q^2]\xi-\Delta\eta+(1-\tilde{\mu}_{\star}^2)r^4 \\ \nonumber
&-2(\sigma Q-\tilde{\mu}_{\star}^2M)r^3+[\sigma^2Q^2+a^2-\tilde{\mu}_{\star}(a^2+Q^2)]r^2  \\ \nonumber
&+2a^2(M-\sigma Q)r-a^2Q^2=0,
\end{align}
\begin{align}
&-(r-M)\xi^2+a(\sigma Q-2M)\xi-(r-M)\eta+2(1-\tilde{\mu}_{\star}^2)r^3 \\ \nonumber
&-3(\sigma Q-\tilde{\mu}_{\star}^2M)r^2+[\sigma^2Q^2+a^2-\tilde{\mu}_{\star}(a^2+Q^2)]r+a^2(M-\sigma Q)=0.
\end{align}
The conserved quantities $\eta=\frac{\mathcal{Q}}{E^2}$ and $\xi=\frac{L_z}{E}$ can be fixed by the radius $r$ of the orbit and the charge parameter $\sigma=\frac{q}{E}$ of the particle. For a null curve, its parameter has a constant-scaling freedom $\lambda\to \alpha^{-1}\lambda$, which induces a scaling transformation for the 4-velocity $U^{\mu}\to\alpha U^{\mu}$. In fact, the equation of motion $U^{\mu}\nabla_{\mu}U^{\rho}=qF^{\rho}_{\phantom{\rho}\mu}U^{\mu}$ has a scaling symmetry defined by $q\to \alpha q$ and $U^{\mu}\to \alpha U^{\mu}$. The energy of the massless particle is $E=g(\partial_t, U)$, so the parameter $\sigma$ is a scaling-invariant ratio. Then from the particle point of view, it is the conserved qualities and $\frac{q}{E}$ that make sense. This is different from that in the QNM analysis, where the frequency, the counterpart of energy, is vital.

Due to the correspondence, we know that $\mathcal{R}(r)=0$ and $\mathcal{R}^{\prime}(r)=0$ corresponds to $V^r(r,\omega)\mid_{r_0,\omega_R}=0$ and $V^r(r,\omega)^\prime\mid_{r_0,\omega_R}=0$ in the QNM analysis, respectively. All the radii of spherical orbits are within $r_1$ to $r_2$, and the inclination angles of these orbits will reach maximal values when $\Theta=0$. We can obtain this extreme by taking $\Theta=0$:
\begin{align}
\cos^2\theta_{\pm}=\frac{a^2(1-\mu^2)-\eta-\xi^2+\sqrt{\left(a^2\left(1-\mu^2\right)-\eta-\xi^2\right)^2+4\eta a^2\left(1-\mu^2\right)}}{2a^2(1-\mu^2)}
\end{align}
into these two equation.

\subsection{Uncharged massless scalar field}
\subsubsection{Redefinition of parameters}
Since the black hole is charged, the gravitational filed around black hole has been changed, so as the motion of particle even if the particle is uncharged. In this section, we will analyze this problem.
In the following two sections we will see the figure of tendency of quasi-normal frequency and the spherical orbit's inclination angle with respect to the parameter of field as well as spacetime. In these pictures we will obtain a more direct understanding about the correspondence. Since there are many parameters in this problem and we are {mainly} interested in the charge effects, we assume in the following discussions that the mass of the scalar field is zero. In order to simplify our discussion, we will redefine some new parameters first.

We define $x=\frac{r}{M}$, $x_0=\frac{r_0}{M}$ (representing the apex of effective potential), $y=M\Omega_{R}$ (representing the real part of frequency), $\lambda=\frac{a}{M}$ (representing the angular momentum of KN black hole), $\Omega=\frac{\omega}{L}$, $k=\frac{Mq}{L}=y\sigma$ (by the correspondence we know $E=\omega_R$), $\rho=\frac{Q}{M}$ and $\mu=\frac{m}{L}$. When we consider the charged massless scalar field, the effective potential of its radial equation is $V^r(r)=\frac{L^2}{M^2}g(x,y)$ with
\begin{align}
g(x,y)=\frac{[y(x^2+\lambda^2)-\mu \lambda-y\sigma\rho x]^2-(x^2-2x+\lambda^2+\rho^2)\left(1+\frac{\lambda^2}{2}\left(1+\mu^2\right)y^2-2\lambda \mu y\right)}{(x^2+\lambda^2)^2}.
\end{align}

If we define
\begin{equation}\label{f1}
f_1=[y(x^2+\lambda^2)-\mu \lambda-y\sigma\rho x]^2-(x^2-2x+\lambda^2+\rho^2)\left(1+\frac{\lambda^2}{2}\left(1+\mu^2\right)y^2-2\lambda \mu y\right)
\end{equation}
and
\begin{equation}\label{f2}
f_2=y[x^2(x-3)+\lambda^2(x+1)+2\rho^2 x]+(\mu \lambda+y\sigma\rho)x-(\lambda^2+\rho^2)y\sigma\rho-\mu \lambda,
\end{equation}
then the conditions $f_1=0$ and $f_2=0$ correspond to the requirements of QNMs, namely $V^r(r,\omega)\mid_{r_0,\omega_R}=0$ and $V^r(r,\omega)^\prime\mid_{r_0,\omega_R}=0$, respectively. We can solve these two equations simultaneously to obtain how $x$ and $y$ depend on $\mu$, $\lambda$, $\rho$ and $\sigma$. We know that $\frac{d}{dr_\star}=\frac{r^2-2Mr+a^2+Q^2}{r^2+a^2}\frac{d}{dr}$ and set $h(x)=\frac{x^2-2x+\lambda^2+\rho^2}{x^2+\lambda^2}$, then the imaginary part of the frequency is
\begin{equation}\label{omegaI}
M\Omega_I(x,y)=\left(n+\frac{1}{2}\right)\frac{\sqrt{2h(x)\frac{d}{dx}[h(x)\frac{d}{dx}g(x,y)]}}{\frac{dg(x,y)}{dy}}.
\end{equation}

\subsubsection{From non-extreme to extreme}
\begin{figure}
\centering
\subfloat[The real part of QN frequency]{
\label{fig:wdfdjsb1}
\includegraphics[width=7.5cm]{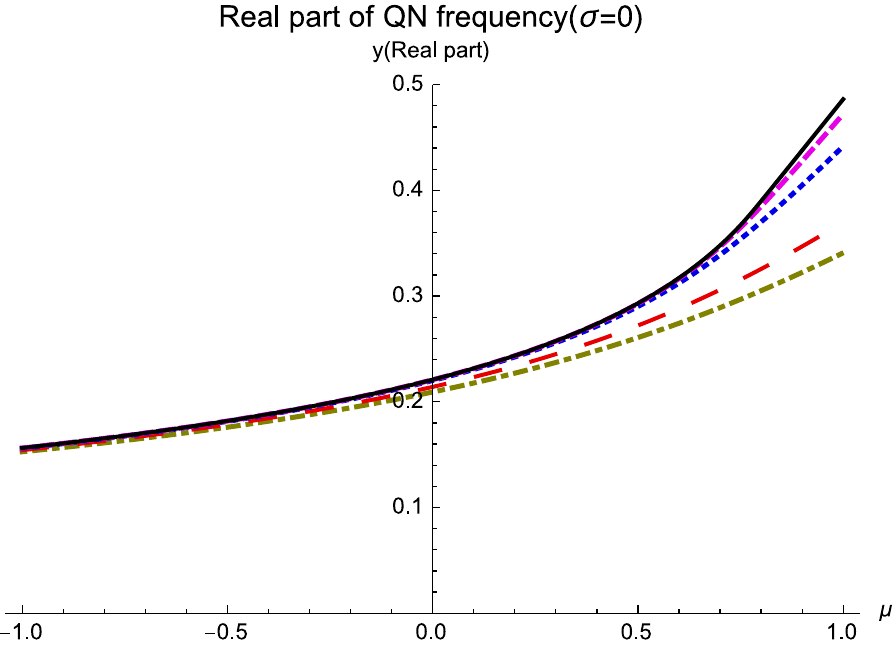}
}
\hspace{8pt}
\subfloat[The imaginary part of QN frequency]{
\label{fig:wdfdjxb1}
\includegraphics[width=7.5cm]{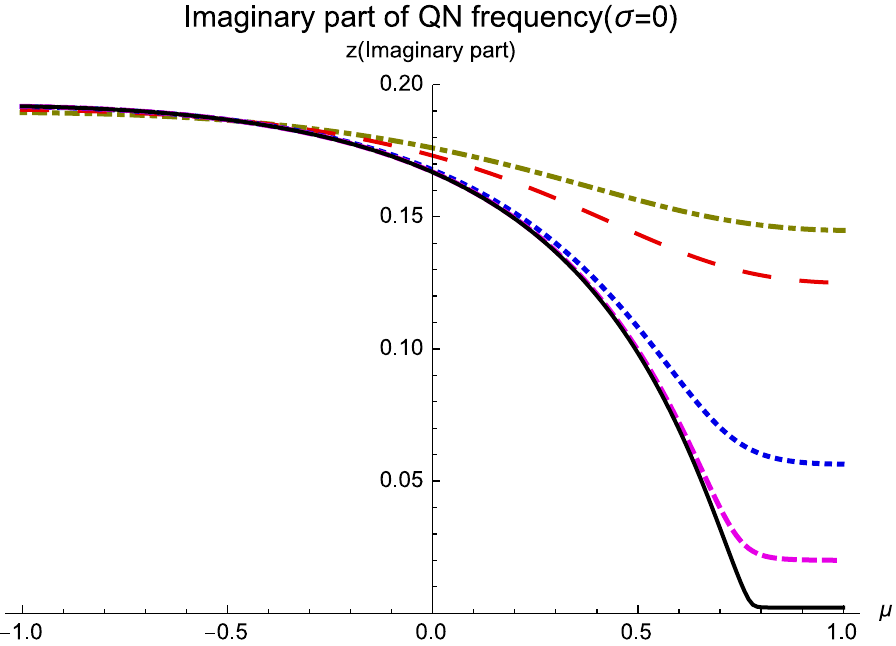}
} \\
\subfloat[The apex of effective potential]{
\label{fig:wdfdjdd11}
\includegraphics[width=7.5cm]{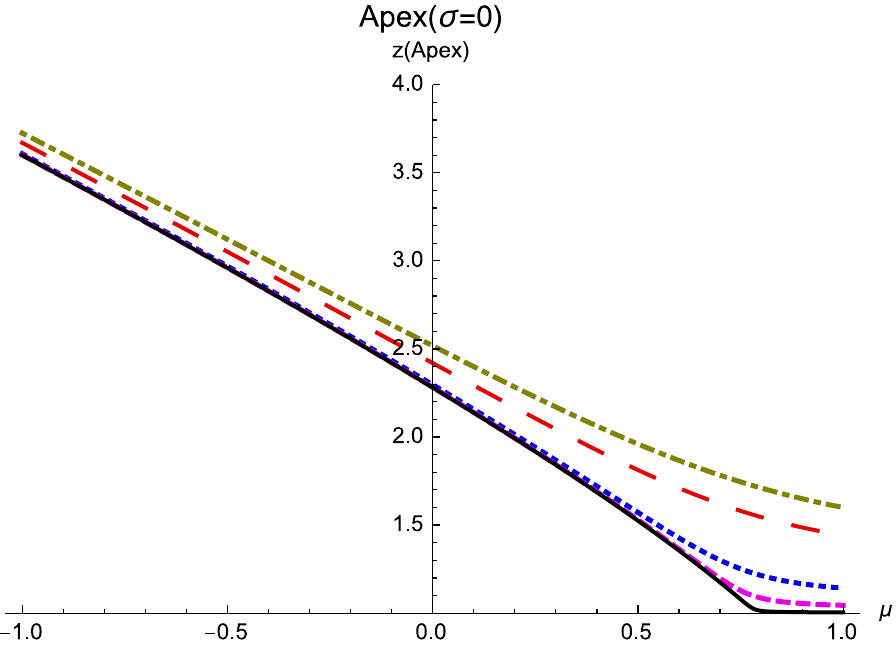}
}
\hspace{10pt}
\subfloat[The inclination angle of spherical orbit]{
\label{fig:wdfdjqjbj1}
\includegraphics[width=7.5cm]{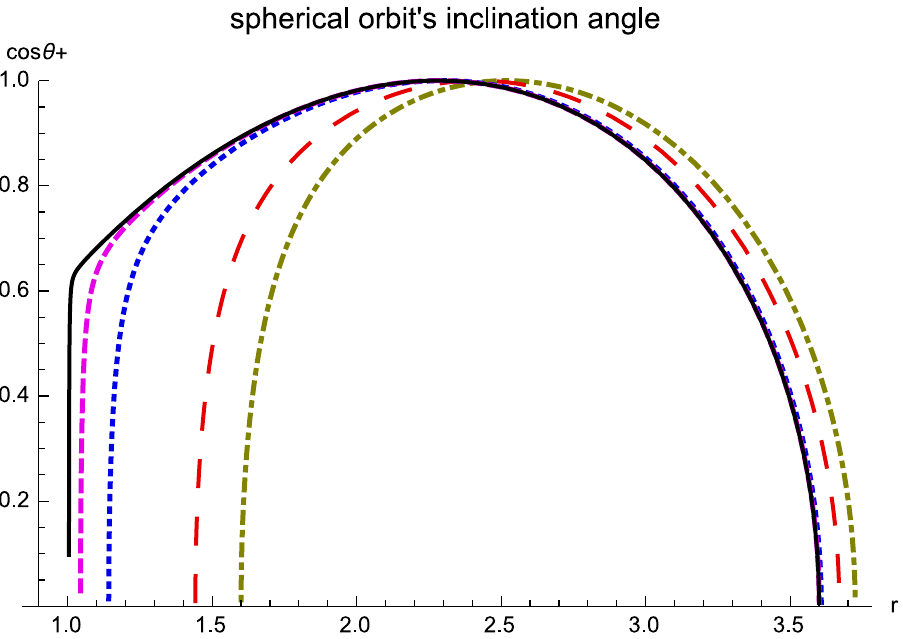}
}
\caption{\footnotesize{From non-extreme to nearly extreme. BH angular momentum $\lambda=\frac{a}{M}=0.8$, the dotdashed dark green line shows $\rho=0.4$ , the large dashed red line shows $\rho=0.5$, the dotted blue line shows $\rho=0.59$, the thick dashed purple line shows $\rho=0.599$, the black line shows $\rho=0.59999$}}
\label{NtEu1}
\end{figure}

Using numerical method, we can solve the equations(\ref{f1}), (\ref{f2})and(\ref{omegaI}), and obtain a series of pictures(Fig.\ref{NtEu1}) about how QNMs and their corresponding orbits changed when KN black hole approach to nearly extreme.

We find that the real part of quasi-normal frequency will increase when the black hole charge increases, which can be found in Fig.\ref{fig:wdfdjsb1}. Furthermore, when the black hole approaches nearly extreme, the relationship between $y$ and $\mu$ approaches linear when $0.788<\mu<1$(strictly speaking, $m=-l...,l$, $\mu = m /(l+1/2)$ never precisely reaches $\pm 1$). What is important is that the imaginary part of quasi-normal frequency approach to zero at $0.788<\mu<1$ when black hole approaches nearly extreme from Fig.\ref{fig:wdfdjsb1}, whose range is different from \cite{H.Yang}. We denote $\mu=0.788$ as $\mu_c$ and we call it turning point. When $\mu>\mu_c$ , the imaginary part of frequency will approach to zero in NEKN(nearly extreme KN spacetime). We can obtain that the apex of effective potential of radial equation of QNMs tends to horizon in NEKN, which can be found in Fig.\ref{fig:wdfdjdd11}. In Fig.\ref{fig:wdfdjqjbj1}, we show the unstable spherical orbit outside the black hole as well as its maximum inclination angle. The horizontal axis tells us the radius range of spherical orbit $r\in$[$r_{min}$,$r_{max}$] . When black hole approaches to extreme we can obtain that many spherical orbits of different maximum inclination angle have nearly the same radius, $r\approx1$, namely the horizon.

Under the geometric-optics correspondence between $(E,L_z,\mathcal{Q})$ and $(\omega,m,A_{lm})$, vanishing imaginary part of quasi-normal frequency corresponds to vanishing Lyapunov exponent of radial motion of particle in NEKN: in the aspect of QNMs, that means the mode will exist slightly outside the horizon for a long time and do not move away very quickly when $\mu>\mu_c$; in the aspect of particle, that means a lot of particle move near the horizon with different maximum inclination angle and do not fall or move away after a perturbation. In addition, Fig.\ref{fig:wdfdjdd11} tells us that the apex $x_0\in[x_{min},x_{max}]$ and  we can see that ($x_{min}$ , $x_{max}$) correspond to ($r_{min}$ , $r_{max}$) of same black hole parameter in  Fig.\ref{fig:wdfdjqjbj1}. That supports the geometric-optics correspondence.

\begin{figure}
\centering
\begin{minipage}[t]{0.4\textwidth}
\centering
\includegraphics[width=7.5cm]{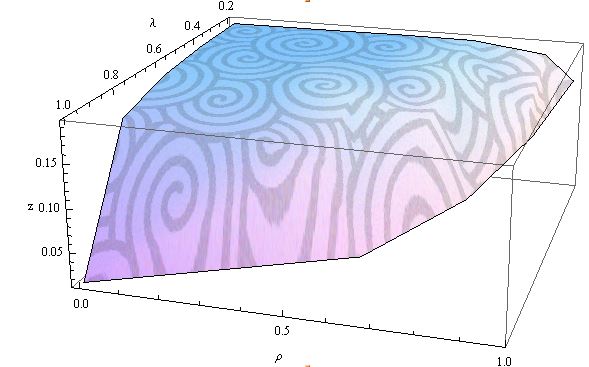}
\caption{\footnotesize{Imaginary part of frequency when $\sigma=0$ and $\mu=0.7$}, $\lambda^2+\rho^2<1$}\label{ucxb}
\end{minipage}
\hspace{80pt}
\begin{minipage}[t]{0.4\textwidth}
\centering
\includegraphics[width=7cm]{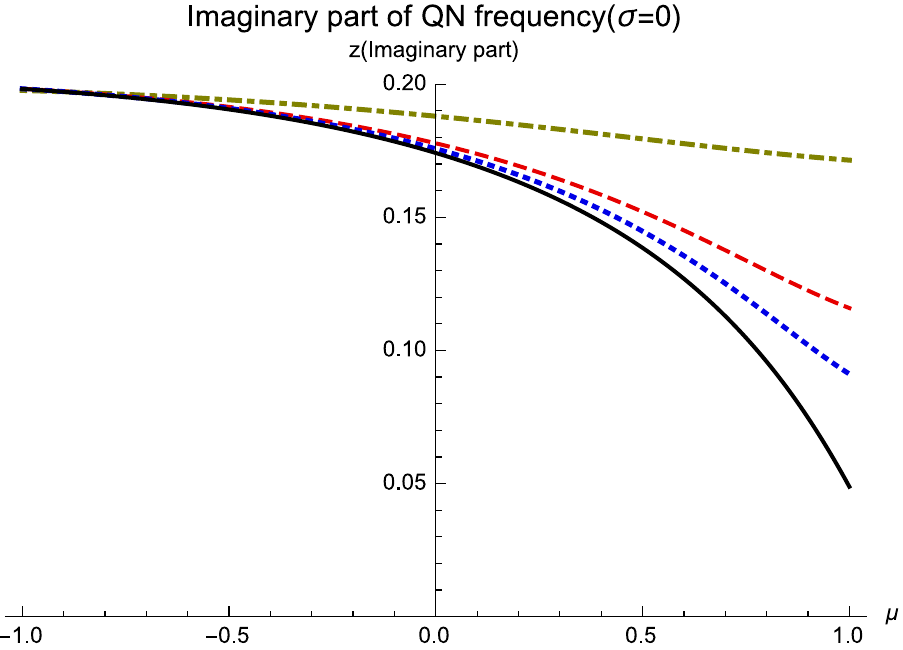}
\caption{\footnotesize{Imaginary part of frequency, the dotdashed dark green line shows
$\rho= 0.8$, the dashed red line shows $\rho=0.9$, the dotted blue line shows $\rho=0.91$, the black line shows $\rho=0.9165$}}\label{wdfdjxb1n}
\end{minipage}
\end{figure}

It is worthy to point out that, unlike Kerr case\cite{H.Yang}, there are infinity ways when KN black hole approaches to nearly extreme. The parameters of KN($\lambda$ and $\rho$) limit in a unit circle($\lambda^2+\rho^2\leq1$), so there are infinity kinds of extreme KN black hole. We show how imaginary part of quasi-normal frequency changes when the parameters of KN change in a unit circle with a fixed $\mu$ in Fig.\ref{ucxb}. We can obtain that the imaginary part approach to zero when $\lambda$ approach to $1$. Furthermore, we can obtain that not every nearly extreme KN black hole has the phenomenon that the imaginary part of QN frequency decay to zero. For example, when $\lambda=0.4$, we can require that the charge of black hole is large enough such that the black hole approach to nearly extreme, we obtain that the imaginary part is always greater than zero from Fig.\ref{wdfdjxb1n}. That is different from \cite{H.Yang}. In next section, we will derive the expression of turning point $\mu_c$ and obtain that $\mu_c$ is large than $1$ in this case.

\subsubsection{The turning point $\mu_c$}
In this section we require that the field is charged but massless. In NEKN, we have $\lambda^2+\rho^2\approx1$, we will solve $f_1=0$ and $f_2=0$ under this condition. $(x=1,y=\frac{\mu \lambda}{1+\lambda^2+\sigma\rho})$ is one of the solutions. $f_2=0$ gives
\begin{equation}
(x-1)[y(x^2-2x-\lambda^2+\sigma\rho)+\mu \lambda]=0.
\end{equation}
Then we can take $y=-\frac{\mu \lambda}{x^2-2x-\lambda^2+\sigma\rho}$ into $f_1=0$ and obtain:
\begin{align}
(\mu\lambda)^2(2x-\rho\sigma)^2=&(x^2-2x-\lambda^2+\sigma\rho)^2+\frac{\lambda^2}{2}(1+\mu^2)(\mu\lambda)^2 \\
&-2(\mu\lambda)^2(x^2-2x-\lambda^2+\sigma\rho).
\end{align}
When considering the turning point of real part of frequency, we take $x=1$ into the equation above and have
\begin{align}
\frac{1}{2}\lambda^4\mu^4-(\frac{3}{2}\lambda^2+\rho^2\sigma^2-6\rho\sigma+6)\lambda^2\mu^2+(\lambda^2-\rho\sigma+1)^2=0
\end{align}
The physically relevant solution of this quartic equation is
\begin{align}\label{chargedmuc}
\mu_c=\frac{1}{\sqrt{2}\lambda}\sqrt{3\lambda^2+12-12\rho\sigma+2\rho^2\sigma^2-\sqrt{B}}
\end{align}
with
\begin{align}
B=136+56\lambda^2+\lambda^4-272\rho\sigma-56\lambda^2\rho\sigma+184\rho^2\sigma^2+12\lambda^2\rho^2\sigma^2-48\rho^3\sigma^3+4\rho^4\sigma^4.
\end{align}

For example, when $k=0$, we have
\begin{equation}\label{chargelessmuc}
\mu_c=\frac{1}{2\lambda}\sqrt{6(\lambda^2+4)-2\sqrt{\lambda^4+56\lambda^2+136}}.
\end{equation}
This is the turning point of QNMs of uncharged and massless scalar field in nearly extreme KN black hole (see \cite{Hod}). This formula (\ref{chargelessmuc}) tells us that $\mu_c$ is less than 1 when $\lambda\in[0.5,1]$, $\mu_c$=0.744 when $\lambda=1$ and $\mu_c=0.788$ when $\lambda=0.8$ which correspond the vanishing imaginary part of frequency when $\mu>\mu_c$. But when $\lambda\in[0,0.5]$ ,the $\mu_c$ will greater than 1, for example $\mu_c$ is $1.17>1$ when $\lambda=0.4$. That corresponds to the fact shows in Fig.\ref{wdfdjxb1n}. This agrees with the discussion in the previous section. So unlike kerr case, when the massless scalar fields are uncharged and the angular momentum $a$ of nearly extreme KN black hole is larger than $0.5M$, the imaginary part of quasi-normal frequency of scalar field will approach to zero with $\mu>\mu_c$.

Another case is that when $\sigma=1$, we can obtain that $\mu_c$ is limited in the range from $0$ to $0.744$ by eq.(\ref{chargedmuc}), so when black hole turn to extreme RN black hole ($\lambda\to0$, $\rho\to1$), the QNMs of charged massless field in RN have zero damping modes (See \cite{Hod2010}).

\subsection{Charged massless field}
In this section we will mainly focus on the effects of field charge.

\subsubsection{Charge effects in non-extreme KN black hole}
\begin{figure}
\centering
\subfloat[The real part of QN frequency]{
\label{fig:tfsb1}
\includegraphics[width=7.5cm]{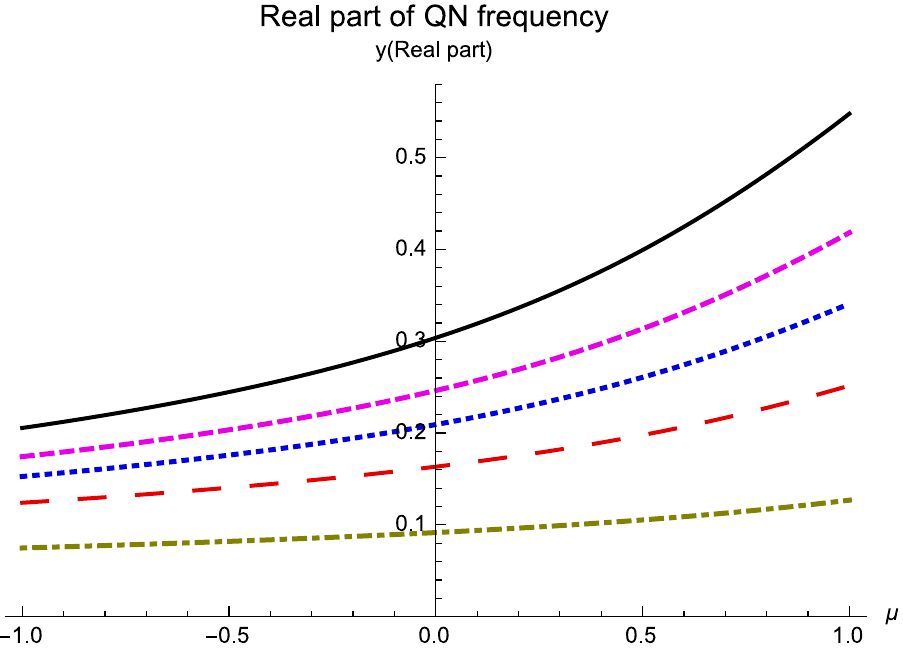}
}
\hspace{8pt}
\subfloat[The imaginary part of QN frequency]{
\label{fig:tfxb1}
\includegraphics[width=7.5cm]{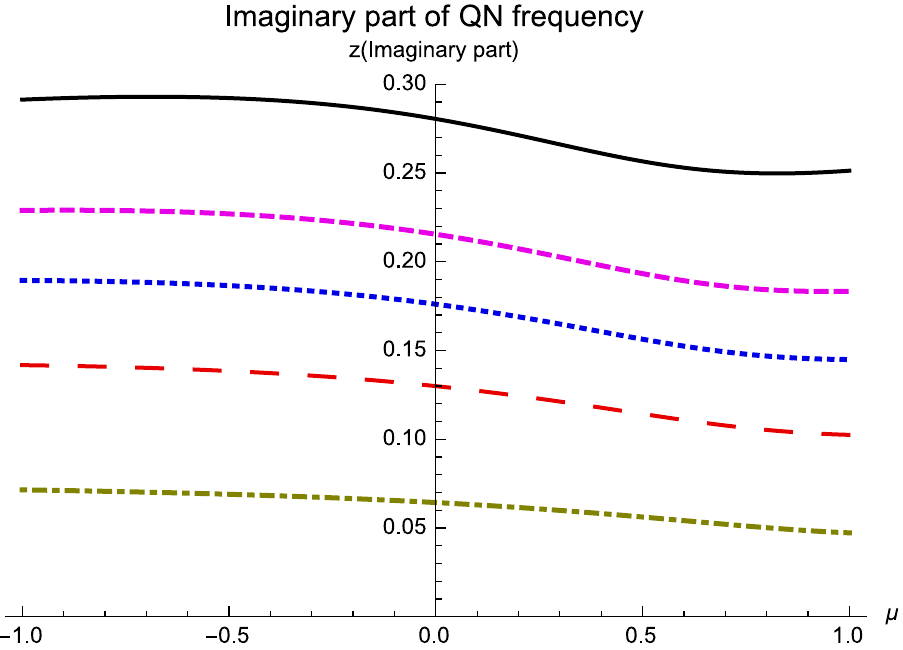}
} \\
\subfloat[The apex of effective potential]{
\label{fig:tfdd1}
\includegraphics[width=7.5cm]{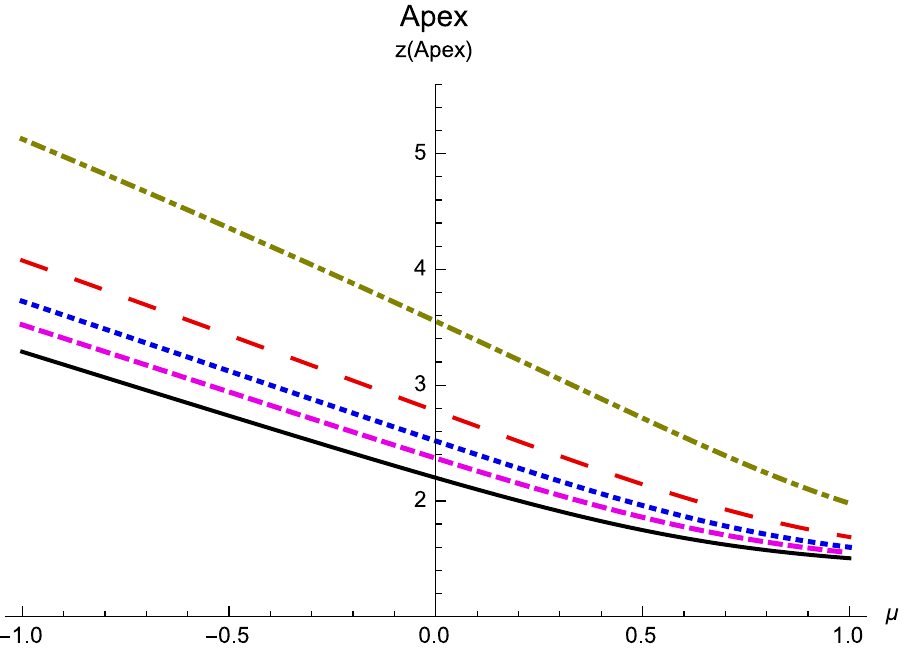}
}
\hspace{10pt}
\subfloat[The inclination angle of spherical orbit]{
\label{fig:fqjbj1}
\includegraphics[width=7.5cm]{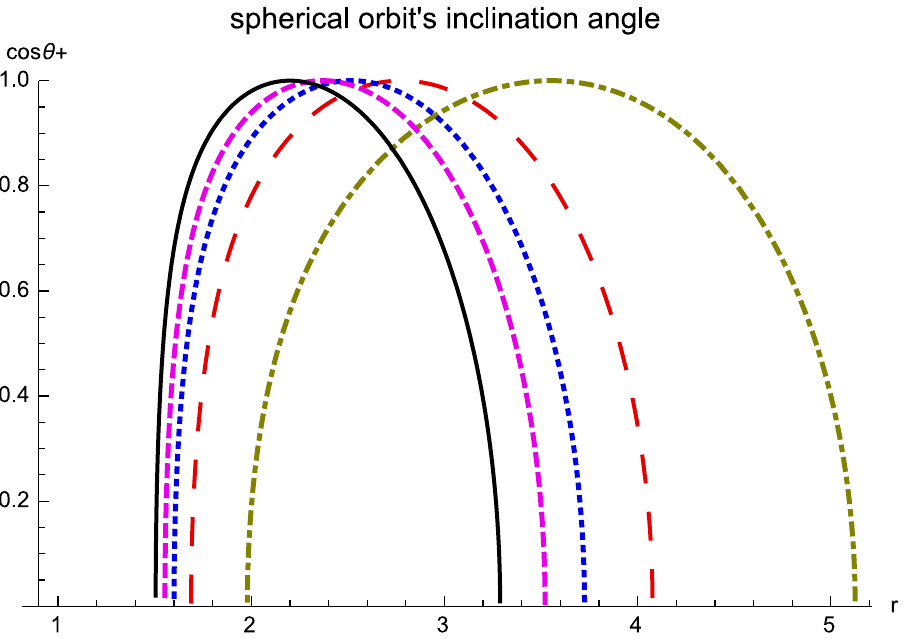}
}
\caption{\footnotesize{The influence of particle charge in non-extreme BH. The BH parameters $\lambda={a}/{M}=0.8$, $\rho=Q/M=0.4$. The dotdashed dark green line represents $\sigma=q/E=-10$, the large dashed red line shows $\sigma=-2$, the thick dotted blue line shows $\sigma=0$, the thick dashed purple line shows $\sigma=1$, and the black line shows $\sigma=2$}.}
\label{neipc}
\end{figure}

Firstly, we want to see the influence in non-extreme KN black hole. We can observe that the real and imaginary parts of the frequency become greater while the apex of effective potential become smaller when the field charge changes from $-10$ to $2$ from Fig.\ref{fig:tfsb1}, Fig.\ref{fig:tfxb1} and Fig.\ref{fig:tfdd1}. When the field charge is opposite to the charge of the black hole, the electromagnetic force is attractive like gravity, these perturbations are located at high orbits and possess relatively smaller energy which contain the gravitational and electromagnetic energy. The electromagnetic forces change from attraction to repulsion when the field charge changes from $-10$ to $2$, then these perturbations need stronger gravity to balance  the electromagnetic force, so they are located at positions closer to the horizon and possess relatively higher energy.

\begin{figure}
\centering
\subfloat[charge~$r$]{
\label{fsigma1}
\includegraphics[width=7.5cm]{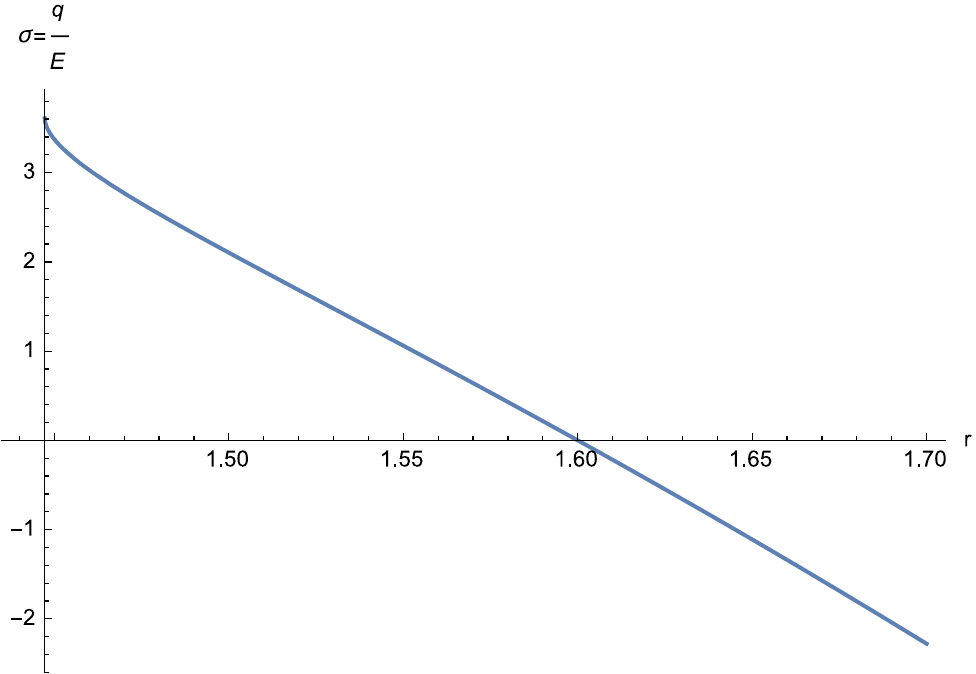}
}
\hspace{8pt}
\subfloat[apex~$\mu$]{
\label{tfdd11}
\includegraphics[width=7.5cm]{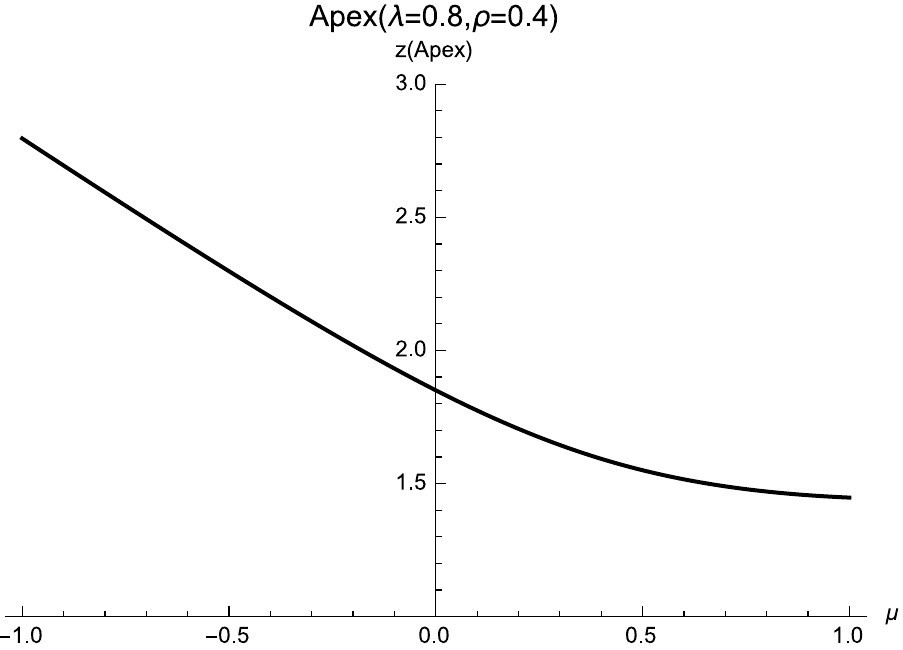}
}
\caption{the range of charge and apex in non-extremal KN BH}
\end{figure}

As we know, the particle orbits that have the maximal and minimal radii locate at the equator, of which the Carter constant is zero. We can rewrite the equation $\mathcal{R}=0$ as
\begin{align}
&r^2-\sigma Qr+a(a-\xi)=\pm(\xi-a)A, \\
&A=\sqrt{r^2-2r+a^2+Q^2}.
\end{align}
Then we substitute this into $\mathcal{R}^\prime=0$ and obtain
\begin{align}
\sigma\big[\pm AQ-\frac{(r-1)Qr}{a\pm A}\big]=\pm2rA-\frac{(r-1)r^2}{a\pm A}.
\end{align}
When we draw the $\sigma$-$r$ picture, we find that we should choose ``$+$''. For example, see Fig.\ref{fsigma1} ($\lambda=a/M=0.8$, $\rho=Q/M=0.4$).

Fig.\ref{fsigma1} shows that the range of $\sigma$ has an upper bound $\frac{r_H}{Q}$, which is saturated as $r\to r_H$. Actually, when we set $\sigma$ to be $\frac{r_H}{Q}$ in this KN black hole, we can draw the apex of the effective potential (Fig.\ref{tfdd11}) in the QNM analysis and obtain that the apex will approaches the horizon when $\mu$ approaches 1.

When the orbit of particle does not totally lying on the equator, or in other words, the Carter constant is not zero, we can not use this method and obtain the range of $\sigma$ easily. In this case, the behaviour of the worldline is much more complicated, and will be considered in future works.
\subsubsection{From non-extreme black hole to extreme black hole}
In this section we analyse the QNMs and their geometric correspondence when black hole approach to extreme.

1. Influence of charge of black hole

\begin{figure}
\centering
\subfloat[The real part of QN frequency]{
\label{fdjsb1}
\includegraphics[width=7.5cm]{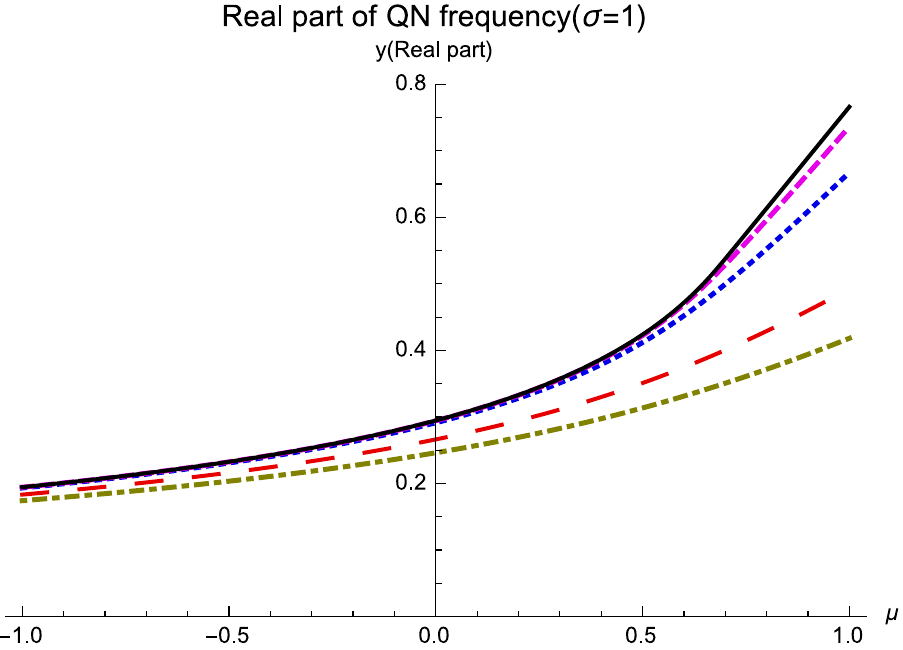}
}
\hspace{8pt}
\subfloat[The imaginary part of QN frequency]{
\label{fdjxb1}
\includegraphics[width=7.5cm]{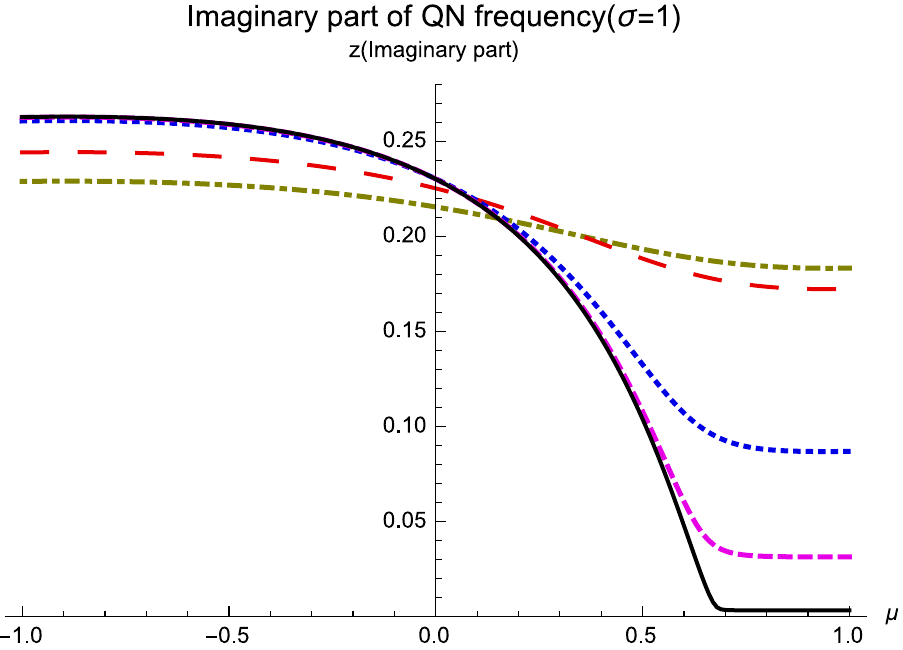}
} \\
\subfloat[The apex of effective potential]{
\label{fdjdd1}
\includegraphics[width=7.5cm]{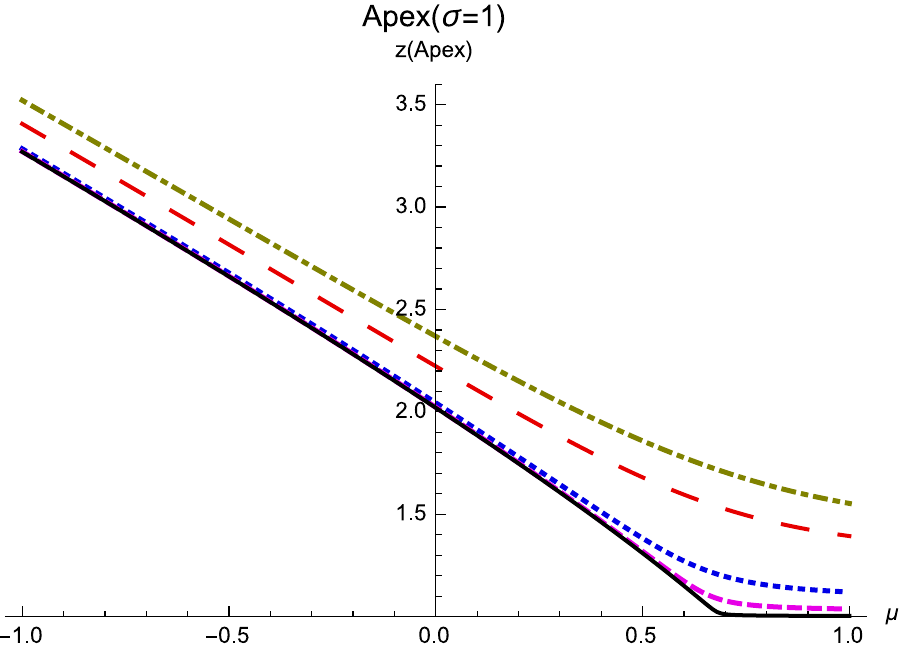}
}
\hspace{10pt}
\subfloat[The inclination angle of spherical orbit]{
\label{fdjqjbj1}
\includegraphics[width=7.5cm]{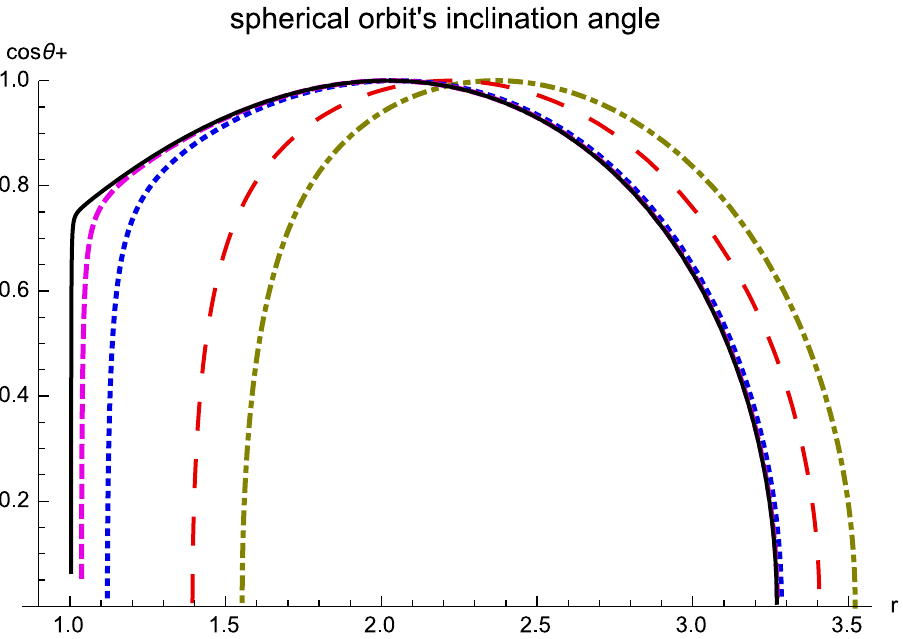}
}
\caption{\footnotesize{The influence of particle charge when KN BH approach to nearly extreme1, with $\sigma=q/E=1$ and the angular momentum of the BH $\lambda=\frac{a}{M}=0.8$. The dotdashed dark green line represents the BH parameters $\rho=Q/M=0.4$, large dashed red line for $\rho=Q/M=0.5$, thick dotted blue line for $\rho=Q/M=0.59$, thick dashed purple line for $\rho=Q/M=0.599$ and black for $\rho=Q/M=0.59999$.}}
\label{fixedmom1}
\end{figure}

We can also obtain the phenomenon of vanishing imaginary part of frequency when black hole approaches to extreme with $\sigma=q/E=1$ and black hole angular momentum $\lambda=a/M=0.8$, which is shown in Fig.\ref{fdjdd1}. (The reasons why we set $\sigma=q/E=1$ and $\lambda=a/M=0.8$ are that we want get the charge effects and there exists zero damping modes in this case.) The phenomenons of real part of frequency, the apex and the inclination angle of spherical orbit are the same like in Fig.\ref{NtEu1}. One of main differents is the position of turning point. From the previous equation of $\mu_c$ (\ref{chargedmuc}) we have derived, we can obtain that $\mu_c=0.688$ when $\lambda=a/M=0.8$ and $\sigma=q/E=1$, Fig.\ref{fdjdd1} also shows us that when $0.688<\mu_c<1$ , the imaginary part of frequency approach zero. The second different is that when the field has the same charge like black hole, the real part of quasi-normal frequency(Fig.\ref{fdjsb1}) will be larger than that of uncharged field(Fig.\ref{fig:wdfdjsb1}) and the apex of effective potential(Fig.\ref{fdjdd1}) is less than that of uncharged field(Fig.\ref{fig:wdfdjdd11}). The explanation is that the electromagnetic force is repulsion and need more gravity to balance. So the charged perturbations will be located near black hole and possess higher total energy than that of uncharged perturbations.

2. Influence of angular momentum of black hole

\begin{figure}
\centering
\subfloat[The real part of QN frequency]{
\label{fdjsb3}
\includegraphics[width=7.5cm]{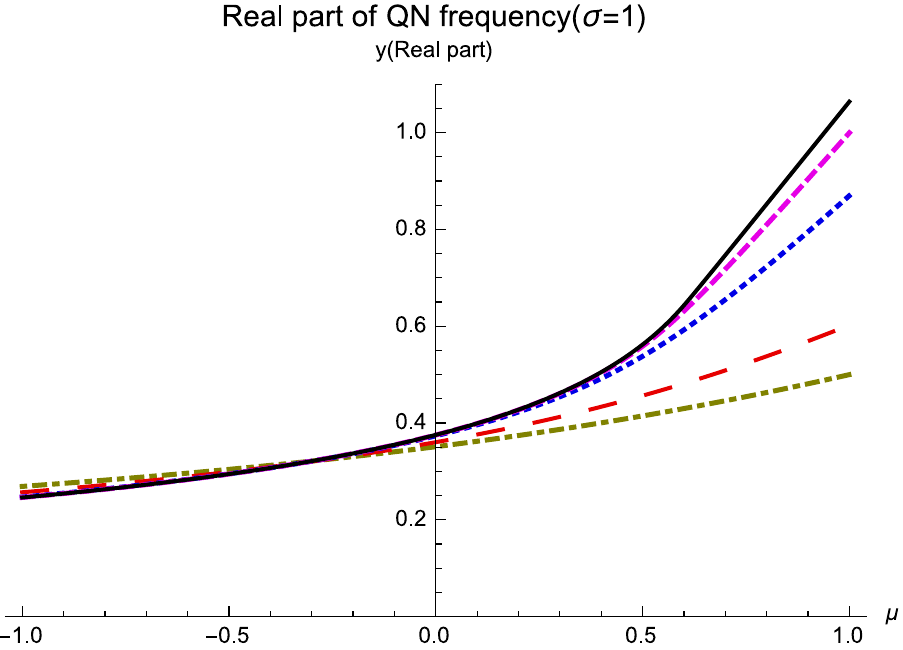}
}
\hspace{8pt}
\subfloat[The imaginary part of QN frequency]{
\label{fdjxb3}
\includegraphics[width=7.5cm]{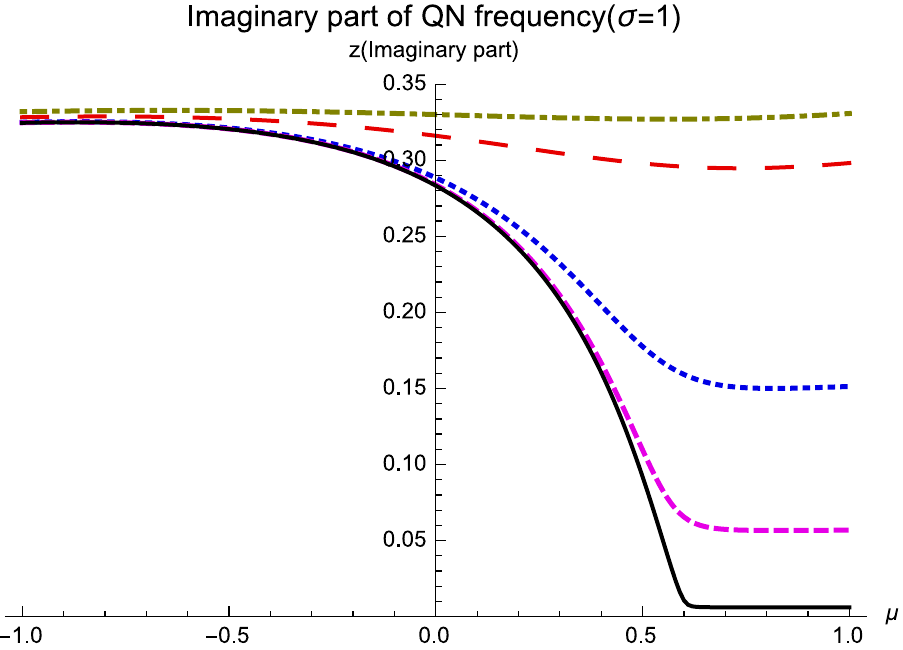}
} \\
\subfloat[The apex of effective potential]{
\label{fdjdd3}
\includegraphics[width=7.5cm]{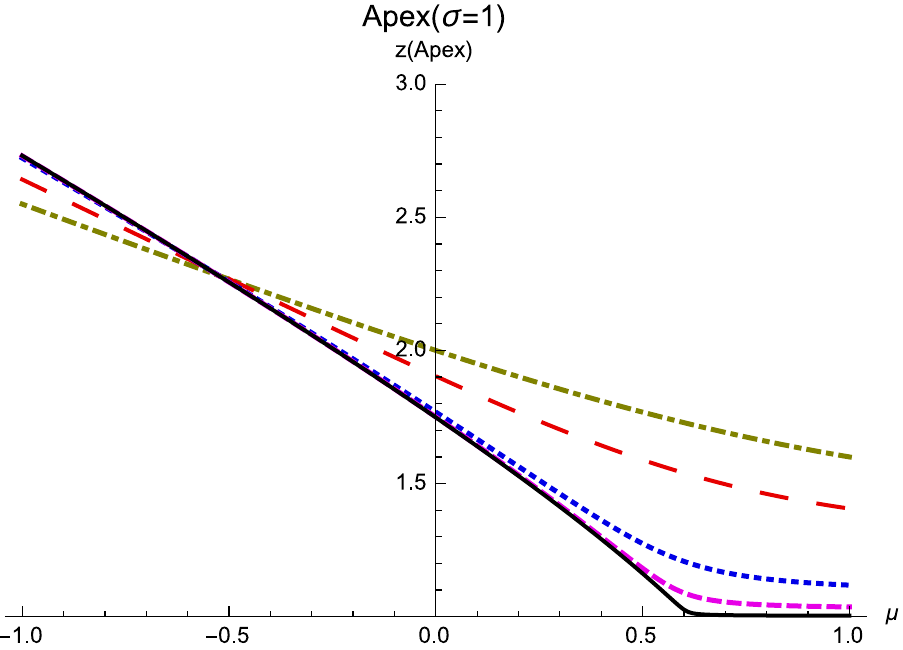}
}
\hspace{10pt}
\subfloat[The inclination angle of spherical orbit]{
\label{fdjqjbj3}
\includegraphics[width=7.5cm]{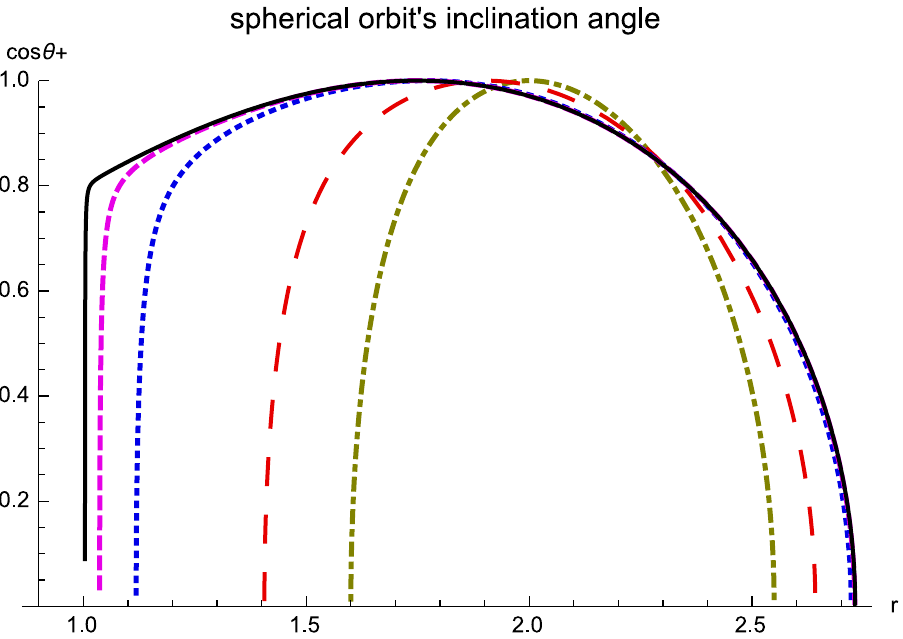}
}
\caption{\footnotesize{The influence of particle charge when KN BH approach to nearly extreme. $\sigma$=q/E=1, BH charge $\rho=Q/M=0.8$, the dotdashed dark green line represent $\lambda=\frac{a}{M}=0.4$, the large dashed red line is 0.5, the thick dotted blue line is 0.59, the thick dashed purple line is 0.599, the black line is 0.59999}}
\label{fixedcharge1}
\end{figure}

When we set a fixed $\sigma=q/E=1$ and black hole charge $\rho=Q/M=0.8$ and let KN black hole approach to nearly extreme(Fig.\ref{fixedcharge1}), we can obtain the phenomenon of vanishing of imaginary part of quasi-normal frequency in NE KN($\rho=0.8,\lambda=0.59999$).
when $0.61<\mu<1$ and $\mu_c=0.61$ from Fig.\ref{fdjxb3}, that correspond to what we have gotten by equation.\ref{chargedmuc}.

3. Disappearance of zero damping modes in NEKN

\begin{figure}
\centering
\begin{minipage}[t]{0.4\textwidth}
\centering
\includegraphics[width=7cm]{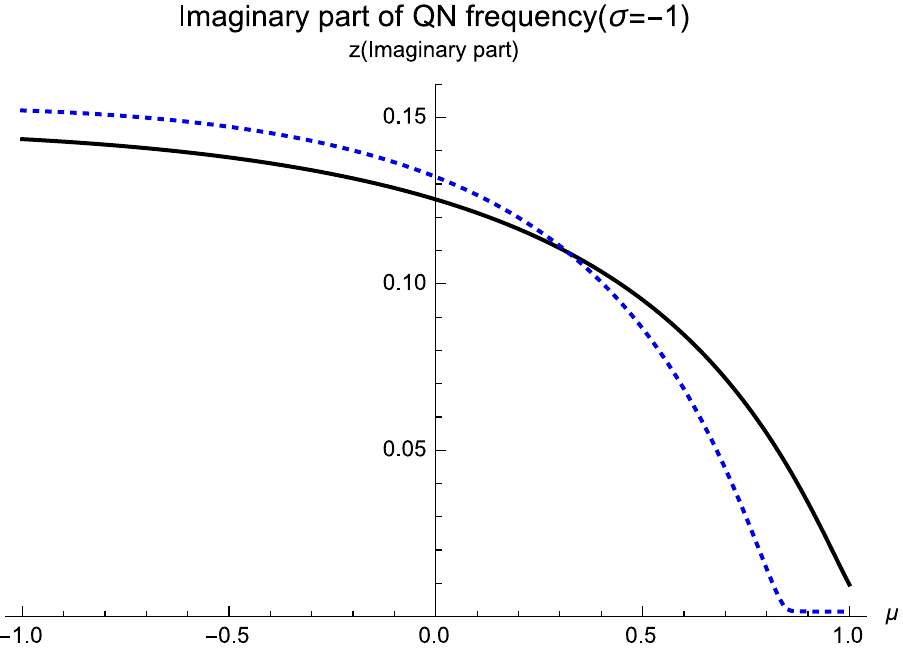}
\caption{\footnotesize{Imaginary part of frequency in NEKN with $\sigma=-1$, the black line shows $\lambda=0.6$ and $\rho=0.79999$, the dotted blue line shows $\lambda=0.8$ and $\rho=0.59999$}}\label{xb4}
\end{minipage}
\hspace{80pt}
\begin{minipage}[t]{0.4\textwidth}
\centering
\includegraphics[width=7.5cm]{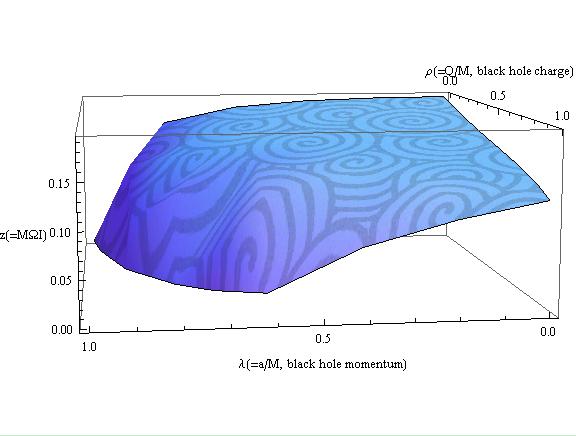}
\caption{\footnotesize{Imaginary part of frequency with $\mu=1$ and $\sigma=-1$}}\label{xb5}
\end{minipage}
\end{figure}

When we use eq.(\ref{chargedmuc}) to calculate the turning point when $\lambda=0.8$, $\rho=0.6$ with $\sigma=-1$, we get $\mu_c=0.857$ , however, when we choose $\lambda=0.6$, $\rho=0.8$ with $\sigma=-1$, we get $\mu_c=1.049>1$. We can obtain that there is no zero damping modes in nearly extreme KN black hole of the last case in Fig.\ref{xb4}.
Considering the variety of NEKN which is different from Kerr case, we show the imaginary part of quasi-normal frequency with fixed $\mu=1$ and $\sigma=-1$ in Fig.\ref{xb5}. We can obtain that there are zero damping modes in a fraction of NEKN, but another fraction of NEKN do not have these modes when $\lambda>0.63$. In the latter case, the angular drag effect of the rotating black hole is not strong enough, so the gravity and the attractive electromagnetic force will pull the particle into black hole, resulting in no circular orbits or zero damping modes.
So the angular momentum of black hole has a more important role to zero damping modes. When the angular momentum is not big enough under extreme case, the imaginary part of quasi-normal frequency will not approach to zero(See \ref{xb5}).

\section{Conclusion}\label{conclusion}
In this paper, we utilize the method of Ref.\cite{H.Yang} to study the QNMs of charged massive scalar field in KN spacetime. We analyse the QNMs by WKB method in eikonal limit($l\gg1$) and get the expression of imaginary part of quasi-normal frequency first. Then we study the Hamilton-Jacobi equations. The charge of particle make the world line congruence be non-geodesic, but we can also find the geometric-optic correspondence under this non-geodesic case and obtain the corresponding relationships between the parameters of QNMs and the conserved qualities of particle in leading order and next-to-leading order by identifying terms in the Hamilton-Jacobi equations and Teukolsky equations. We use these results to study the influence of charge in details. We can observe the phenomenon of vanishing imaginary part of quasi-normal frequency in extreme KN black hole by numerical method and then we obtain the expression of turning point by analytical method. The expression of turning point which depends on the parameters of black hole as well as charge of field gives us the range of zero damping modes. We also obtain a rough range of the charge of field about when QNMs can exist in KN black hole.

Under the geometric-optics correspondence, the massive scalar field should correspond to timelike worldlines, instead of null ones, so the behaviour will be more complicated. Furthermore, in the case with the massive scalar field, the black hole will in general develops a super-radiant instability (see, e.g., \cite{Nambo,Cardoso2015}), but it is not yet clear how (if possible) to see signals of such instability in the WKB approximation.\footnote{We thank the referee for pointing out this issue.} This is also an interesting topic and should be considered in future.

\section{Acknowledge}
We thank Hongbao Zhang for useful discussions and comments. This work is partly supported by the National Natural Science Foundation of China (Grant Nos. 11175245 and 11475179).


\begin{thebibliography}{99}

\bibitem{C.V.Vishveshwara} C. V. Vishveshwara, Nature {\bf 227}, 936 (1970).

\bibitem{Press}  W. H. Press, Long wave trains of gravitational waves from a vibrating black hole, Astrophys. J. 170, 105-108 (1971).

\bibitem{ChandraDetweiler1975} S. Chandrasekhar and S. Detweiler, Proc.
R. Soc. Lond. A {\bf 344}, 441 (1975).

\bibitem{Leaver} E. W. Leaver, Proc. R. Soc. Lond. A {\bf 402}, 285
(1985).

\bibitem{Hod2010} S. Hod, Phys. Lett. A 374,2901(2010).

\bibitem{Hod2012} S. Hod, Phys. Lett. B 710,349(2012).

\bibitem{Hod2015} S. Hod, Phys. Lett. B 747,339(2015).

\bibitem{Richartz} M. Richartz and D. Giugno, Phys. Rev. D. 90, 124011 (2014).

\bibitem{Ferrari2008} V. Ferrari and L. Gualtieri, Gen. Relativ. Gravit.
{\bf 40}, 945 (2008).

\bibitem{Kokkotas} K. D. Kokkotas and B. G. Schmidt, Quasi-Normal Modes of Stars and Black Holes, Living Rev. Rel. 2, 2 (1999).

\bibitem{Berti2009} E. Berti, V. Cardoso and A. O. Starinets, Class. Quant. Grav. 26, 163001 (2009).

\bibitem{Konoplya} R. A. Konoplya and A. Zhidenko, Rev. Mod. Phys.
{\bf 83}, 793 (2011).

\bibitem{Ferrari1984} V. Ferrari and B. Mashhoon, Phys. Rev. D {\bf 30},
295 (1984).

\bibitem{schutz} B. F. Schutz and C. M. Will, Astrophys. J. {\bf 291},
L33 (1985).

\bibitem{Cardoso2009} V. Cardoso, A. S. Miranda, E. Berti, H. Witek and V. T. Zanchin, Phys. Rev. D 79, 064016 (2009).

\bibitem{H.Yang} H. Yang, D. A. Nichois, F. Zhang, A. Zimmerman, Z. Zhang and Y. Chen, Phys. Rev. D 86, 104006 (2012).

\bibitem{J.D.Bekenstein} J. D. Bekenstein, Hydrostatic equilibrium and gravitational collapse of relativistic charged fluid balls, Phys. Rev. D 4, 2185 (1971).

\bibitem{E.Olson} E. Olson and M. Bailyn, Charge effects in a static, spherically symmetric, gravitating fluid, Phys. Rev. D 13, 2204 (1976).

\bibitem{Howell} D. A. Howell et al., The type Ia supernova SNLS-03D3bb from a super-chandrasekhar-mass white dwarf star, Nature (London) 443, 308 (2006).

\bibitem{Scalzo} R. A. Scalzo et al., Nearby supernova factory observations of SN 2007if: First total mass measurement of a super-Chandrasekhar-mass progenitor, Astrophys. J. 713, 1073 (2010).

\bibitem{Olson} E. Olson and M. Bailyn, Charge effects in a static, spherically symmetric, gravitating fluid, Phys. Rev. D 13, 2204 (1976).

\bibitem{Herrera} L. Herrera and W. Barreto, Newtonian polytropes for anisotropic matter: General framework and applications, Phys. Rev. D 87, 087303 (2013).

\bibitem{Xiangdong Zhang} H. Liu, X. Zhang and D. Wen, Phys. Rev. D 89, 104043 (2014).

\bibitem{Berti} E. Berti and K. D. Kokkotas, Phys. Rev. D 71 (2005) 124008.

\bibitem{Hod2008} S. Hod, Phys. Lett. B 666, 483-485 (2008).

\bibitem{Kokkotas2011} K. D. Kokkotas, R. A. Konoplya and A. Zhidenko, Phys. Rev. D 83, 024031 (2011).

\bibitem{Konoplya2013} R. A. Konoplya and A. Zhidenko, Phys. Rev. D 88, 024054 (2013).

\bibitem{Mark2014} Z. Mark, H. Yang, A. Zimmerman and Y. Chen, Phys. Rev. D 91, 044025 (2015).

\bibitem{Hackmann2013} E. Hackmann and H. Xu, Phys. Rev. D 87, 124030 (2013).

\bibitem{Hackmann2015} E. Hackmann and C. Lammerzahl, [arxiv:1506.01572].

\bibitem{Ulbricht} S. Ulbricht and R. Meinel, [arxiv:1503.01973].

\bibitem{Teukolsky} S. A. Teukolsky, Phys. Rev. Lett. {\bf 29}, 1114
(1972).

\bibitem{Chandrasekhar} S. Chandrasekhar, {\it The Mathematical Theory of Black Holes} (Oxford University Press, Oxford, 1983).

\bibitem{Hod} S. Hod, Phys. Lett. B 715, 348 (2012).

\bibitem{Nambo} H. Furuhashi and Y. Nambu, Prog. Theor. Phys. 112 (2004) 983-995. 

\bibitem{Cardoso2015} R. Brito, V. Cardoso and P. Pani, [arXiv:1501.06570].


\end{thebibliography}
\end{document}